\documentclass[twocolumn]{aastex62}

\usepackage{amsmath}
\usepackage{amssymb}

\graphicspath{{./}{figures/}}

\received{January 1, 2018}
\revised{January 7, 2018}
\accepted{\today}
\submitjournal{ApJ}

\shortauthors{Telikova et al.}

\newcommand{\HI}{H\,{\sc i}}
\newcommand{\HII}{H\,{\sc ii}}
\newcommand{\HeI}{He\,{\sc i}}
\newcommand{\HeII}{He\,{\sc ii}}

\newcommand{\CIV}{C\,{\sc iv}}
\newcommand{\MgII}{Mg\,{\sc ii}}

\newcommand{\SiII}{Si\,{\sc ii}}
\newcommand{\SiIII}{Si\,{\sc iii}}
\newcommand{\SiIV}{Si\,{\sc iv}}
\newcommand{\OI}{O\,{\sc i}}

\usepackage{float}
\usepackage[caption = false]{subfig}

\begin{document}

\title{{Thermal state} of the intergalactic medium at $z\sim2-4$}

\correspondingauthor{K.~N.~Telikova}
\email{telikova.astro@mail.ioffe.ru}

\author[0000-0002-7919-245X]{K.~N.~Telikova}
\author[0000-0002-5810-668X]{P.~S.~Shternin}
\author[0000-0002-3814-9666]{S.~A.~Balashev}
\affiliation{Ioffe Institute, Politekhnicheskaya  
 26, St.~Petersburg, 194021, Russia}

\turnoffedit

\begin{abstract}

We present a new method to infer parameters of the temperature-density relation in the intergalactic medium in the post-reionization epoch at $z\sim 2-4$. This method is based on the analysis of the Ly$\alpha$ absorbers distribution over column densities and Doppler parameters by the model joint probability density function. This approach allows us to measure the power-law index $\gamma$ of the temperature-density relation and a certain combination of the temperature at the mean density $T_0$ and hydrogen photoionization rate $\Gamma$.
To estimate $T_0$ and $\Gamma$ separately, we employ measurements of the Ly$\alpha$ forest effective opacity and the model gas probability density function. We show that $\gamma$ tends to be lower than 1.6 and reaches 1.3 at redshift $\sim3$. The inferred temperatures at the mean density are $\sim(2\pm0.5)\times10^4$\,K in the studied redshift range. Both these estimates favour \HeII\ reionization at $z\gtrsim3$. We find that the hydrogen photoionization rate is $\sim0.6\times10^{-12}$\,s$^{-1}$, which is consistent with previous measurements.  

\end{abstract}


\section{Introduction} \label{sec:intro}
It is well established that in the $\Lambda$CDM cosmological model most baryons reside in the intergalactic medium (IGM), especially at high redshifts ($z\sim3$) \citep{Rauch1998, McQuinn2016}.
The baryons in the IGM trace the dark matter dynamical evolution until the gas pressure becomes important. 
While the density evolution of the IGM at moderate overdensity as well as {the evolution} of the dark matter are well described following the Zeldovich approximation \citep{Zeldovich1970,HuiGnedin}, the IGM thermal state is determined by the ionizing background radiation ubiquitous after the birth of the first stars. 

Despite the considerable progress in the IGM studies, our understanding of the thermal history of the Universe is far from complete. There are two key reionization events driven by \HI\ (as well as \HeI) and \HeII\ ionization, each of them leading to changes in the IGM {temperature-density relation (TDR)}. Redshifts and durations of the reionization events, properties and nature of the UV background sources 
{strongly influence}  the IGM {TDR}. Therefore the measurements of the latter can be used to constrain the reionization history and reconstruct the ionizing UV background and its evolution.

Quasars provide a unique opportunity to probe physical conditions in the IGM. Intergalactic clouds imprint themselves in the form of a so-called Ly$\alpha$ forest in spectra of bright background sources. The Ly$\alpha$ forest contains information on a small-scale structure and a thermal evolution of the IGM. High-resolution quasar spectra obtained with advanced optical telescopes such as the Keck Telescope and Very Large Telescope allow to infer this information. 
Several methods to constrain physical properties of the IGM, particularly the {TDR}, were developed in the past based on various properties of the Ly$\alpha$ forest in quasar spectra. Among them are Ly$\alpha$ forest flux probability density analysis \citep[e.g.][]{Lee2015}, its power spectrum analysis \citep[e.g.][]{Rorai2017a,Zaldarriaga2001}, wavelet decomposition of the Ly$\alpha$ forest \citep[e.g.][]{Lidz2010,Garzilli2012}, curvature statistics \citep[e.g.][]{Becker2011}, etc. 
The results obtained by these methods have not converged yet to the complete agreement. 
For instance, there is no consensus whether there are epochs where the IGM is described by an inverted temperature-density relation or not.

More direct method is the statistical analysis of the parameters of the IGM clouds obtained from the Voigt profile fitting of the Ly$\alpha$ forest lines
\citep[e.g.][]{Schaye1999,Schaye2000,Rudie2012,Hiss2017,Hiss2019}. 
Unfortunately, this method has difficulties, because the multicomponent Voigt profile fitting procedure
can sometimes lead to ambiguous results since it strongly relies on a choice of a velocity structure of the absorber.
Moreover, the Ly$\alpha$ forest lines can be contaminated by metal absorption features, located by chance in the same wavelength region. 

In this paper we further develop the method of estimation of the IGM {TDR} using the Voigt profile fitting of the solitary Ly$\alpha$ lines. Via the original procedure \citep{Telikova2018a,Telikova2018b}, we obtained a sample of the solitary Ly$\alpha$ forest lines from  47 high-resolution quasar spectra within a wide redshift range $z=2-4$. Approximation of the obtained distribution of the Ly$\alpha$ forest lines over the column densities $N$ and Doppler parameters $b$ by a model joint probability density function allowed us to estimate the power-law index $\gamma-1$ of the {TDR} and a combination of the temperature at the mean density $T_0$ and the hydrogen photoionization rate $\Gamma$.  Then, using the measurements of the effective optical depth from \citet{Faucher2008tau} {and the model gas density probability distribution function} we independently inferred $T_0$ and $\Gamma$ parameters.

The paper is organized as follows. In Section~\ref{sec:formalism} we present the basic assumptions and formalism used to estimate {TDR} parameters. In Section~\ref{sec:data} we describe the quasar data sample used in this work. We present the procedure for quasar spectra analysis in Section~\ref{sec:fitting procedure}. Metal line rejection and the analysis of the Ly$\alpha$ forest sample by the model probability density function is presented in Section~\ref{sec:analysis of the b-N sample}. 
We discuss the results in Section~\ref{sec:discussion} and summarize in Section~\ref{sec:conclusions}.

In what follows, we assume a standard $\Lambda$CDM cosmology with the matter, dark energy, and baryon density parameters $\Omega_m = 0.28$, $\Omega_\Lambda = 0.72$, and $\Omega_b = 0.046$, respectively, and with the reduced Hubble constant $h=0.70$. We also assume that helium is fully ionized at the probed redshifts.

\section{Basic equations} 
\label{sec:formalism} 

\subsection{Voigt profile}
\label{sec:voigt}

Ly$\alpha$ forest absorption lines (and other spectral lines as well) in quasar spectra are usually described by the Voigt profile. In this approach, the profile of an absorption line is determined 
by the optical depth, which, in the rest frame of an absorber, can be written as
\begin{equation}
	\label{eq:tau}
	\tau(\lambda) = \sqrt{\pi}\frac{e^2}{m_ec} \frac{\lambda_{ik}f_{ik} N}{b} H(a,\,x) \equiv \tau_0 H(a,\, x),
\end{equation}
where $e$ and $m_e$ are a charge and a mass of an electron, respectively, $c$ is the speed of light, $\lambda_{ik}$ and $f_{ik}$ are, respectively, the wavelength and oscillator strength of the transition. The parameter $b$ in Equation~(\ref{eq:tau}) is the width of the projected velocity distribution and $N$ is the column density of absorbing atoms. $H(a,\,x)$ is a Voigt function
\begin{equation}
	\label{eq:voigt}
	H(a,x) = \frac{a}{\pi} \int\limits_{-\infty}^{+\infty} \frac{e^{-y^2}}{(x-y)^2+a^2} dy,
\end{equation}
where $x = (\lambda - \lambda_{ik})/\Delta\lambda_D$, $\Delta\lambda_D =\lambda_{ik} b/c$, $a = \gamma_{ik} \lambda_{ik}/(4 \pi b)$, and $\gamma_{ik}$ is the damping constant. The observed lines are shifted and broadened due to the cosmological expansion as described
 by the cosmological redshift $z$ 
by making a substitution $\lambda \to \lambda / (1+z)$ in Equation~(\ref{eq:tau}).

The Voigt function in Equation~\eqref{eq:voigt} arises from a convolution of the Gaussian and Lorentzian functions, where the Lorentzian function describes a cross section of bound-bound transitions and the Gaussian function describes a hydrogen atoms velocity distribution projected on a line of sight. 
The Gaussian form for the velocity distribution is appropriate when the motions are thermal. It is also a good approximation for the turbulent motions under the microturbulence assumption.
In fact, for \HI\ lines with Doppler parameters $b\sim(20-60)$\,km~s$^{-1}$ (typical for the Ly$\alpha$ forest) and column densities 
$N\lesssim 10^{16}$~cm$^{-2}$, the Voigt function can be reduced to the simple Gaussian function. 

In summary, for the transition with known atomic parameters ($\lambda_{ik}$, $f_{ik}$, and $\gamma_{ik}$), the line profile is described only by three parameters: the width of the projected velocity distribution $b$, the column density of absorbing atoms $N$, and the redshift $z$ of the absorption system.

The actual intensity observed in the spectrum is determined by a convolution of the unabsorbed background source continuum, $I_c(\lambda)$, multiplied by the absorption line profile, with an instrument response function $G$,
\begin{equation}
	\label{eq:convol}
	I_{\rm conv}(\lambda) = \int\limits_{-\infty}^{+\infty} I_c(\lambda') e^{-\tau(\lambda')} G(\lambda,\,\lambda')  d\lambda'.
\end{equation}

In case of optical spectra, the response function is usually symmetric, and weakly depends on the position in the spectra, i.e. $G(\lambda,\,\lambda')=G(\lambda-\lambda')$, and can be well approximated by the Gaussian, with a width corresponding to the spectral resolution. 

\subsection{IGM TDR}
\label{sec:igm_eos}

It was shown by \cite{HuiGnedin} that a temperature-density relation, $T(\rho)$, in the low-density IGM after the \HI\ reionization obeys the power-law form
\begin{equation}\label{eq:eos}
  T=T_0 \left( \frac{\rho}{\bar{\rho}} \right) ^{\gamma-1}\equiv T_0 \Delta^{\gamma-1},
\end{equation}
where $\Delta = \rho/\bar\rho$ is a mass overdensity, $\bar \rho$ is the mean 
density of the Universe, and $T_0$ is the temperature at the mean density. 
Far from reionization events, a power-law index $\gamma-1$ of the {TDR} is approximately 0.6 \citep{HuiGnedin}, while close to the reionization it can be significantly lower. 

The IGM {TDR manifests} itself in the distribution of the Ly$\alpha$ forest absorbers in the $N-b$ parameter space as a sharp density contrast at low values of $b$ parameters. This contrast is called a cutoff of the distribution and it is thought 
to be caused by the pure thermal broadening with negligible contribution of peculiar motions in the cloud \citep{Schaye1999}. In this case, the position of the cutoff $b(N)=b_\mathrm{th}(N)$ is
\begin{equation}\label{min_b}
b_{\rm th} = \sqrt{2k_{\rm B}T/m},
\end{equation}
where $k_{\rm B}$ is the Boltzmann constant and  $m$ is the hydrogen atom mass. Under the assumptions of the uniform UV background and the characteristic size of the absorber being equal to the Jeans length, the column density of neutral hydrogen is, in turn, related to the IGM parameters as \citep{Schaye2001} 

\begin{equation}\label{eq:rho-N}
	N = 1.7\times 10^{13}\, \frac{\Delta^{3/2}}{\Gamma_{-12}}\left(\frac{T}{10^4~\mathrm{K}}\right)^{-0.22}\left(\frac{1+z}{3.4}\right)^{9/2}~{\rm cm}^{-2},
\end{equation}
where $\Gamma_{-12}$ is a hydrogen photoionization rate $\Gamma$ in $10^{-12}$~s$^{-1}$ units. It depends on a spectrum 
of the UV ionization background and is usually inferred either from the Ly$\alpha$ forest opacity measurements using predefined IGM {TDR} parameters or calculated using the specified model of the UV background.
In deriving Equation~(\ref{eq:rho-N}) and below, we use the hydrogen recombination rate  $R(T)\approx4\times 10^{13} \left(T/(10^4~ \rm{K})\right)^{-0.72} $~cm$^3$~s$^{-1}$  \citep{Draine2011}.

According to Equations~(\ref{eq:eos})--({\ref{eq:rho-N}}), the relation between the thermal broadening in the absorber and its column density has a power-law form
\begin{equation}\label{eq:b-N}
 b_{\rm{th}}= b_0\left(\frac{N}{10^{12}~{\rm cm}^{-2}}\right)^{\xi-1} \left(\frac{1+z}{3.4}\right)^{-9(\xi-1)/2}, 
\end{equation}
where $b_0$ is a normalization constant,

\begin{equation}\label{eq:b_0}
b_0 = 12.8  \left(\frac{\Gamma_{-12}}{17}\right)^{\xi-1} \left(\frac{T_0}{10^4\ \mathrm{K}}\right)^{1/2+0.22(\xi-1)} \mathrm{km}\ \mathrm{s}^{-1}
\end{equation}
and the index $\xi$ is related to the power-law index $\gamma$ in Equation~(\ref{eq:eos}) as
\begin{equation}\label{eq:Gamma-gamma}
    \xi-1 = \frac{\gamma-1}{3-0.44(\gamma-1)}.
\end{equation}

Thus, if the cutoff position in the $b-N$ plane (at different redshifts) is determined,  Equations~(\ref{eq:rho-N})--(\ref{eq:Gamma-gamma}) allow one to probe the IGM {thermal state} and constrain $\gamma(z)$, $T_0(z)$, and $\Gamma(z)$. 
Notice, that the cutoff slope $\xi$ is a robust measure of the power-law index $\gamma$ via Equation~(\ref{eq:Gamma-gamma}) which does not depend on the exact coefficient in Equation~(\ref{eq:rho-N}). In contrast, the value of this coefficient (determined by the actual relation between absorber size and Jeans length) can make the determination of the parameters $T_0$ and $\Gamma$ from the cutoff intercept $b_0$ somewhat ambiguous. 
Moreover,  Equation~(\ref{eq:b_0}) allows to constrain only a certain combination of $T_0$ and $\Gamma$. This degeneracy
should be removed with additional constraints.
A reasonable independent constraint on the parameters $T_0(z)$ and $\Gamma(z)$ employs measurements of an effective Ly$\alpha$ forest opacity, which corresponds to the optical depth of Ly$\alpha$ transitions averaged over many sightlines.

The effective optical depth in question can be expressed via a local optical depth $\tau$ and a gas probability density distribution $P(\Delta,z)$ as \citep{Faucher2008}
\begin{equation}\label{eq:tau_eff}
    \tau_{\rm eff}(z) = -\ln\left[\int^\infty_0 P(\Delta,\,z)e^{-\tau(z)} {\rm d}\Delta\right].
\end{equation}
Here the local optical depth is taken in the Gunn-Peterson approximation \citep{Gunn&Peterson1965ApJ}
\begin{equation}\label{eq:tau_Gunn-Peterson}
    \tau = \frac{\pi e^2 f}{m_e \nu}\frac{1}{H(z)}\frac{R(T)}{\Gamma}n_{\rm \text{\HII}}n_e, 
\end{equation}
where $\nu$ is the frequency and $f$ is the oscillator strength of the hydrogen Ly$\alpha$ transition, $n_{\rm \text{\HII}}$ and $n_e$ are \HII\ and electrons number densities, respectively, $H(z)$ is the Hubble constant. In this approach, individual density profiles are averaged, and the absorber structure and broadening of associated lines are smoothed out. However, the incidence rate of Ly$\alpha$ forest lines and the shape of the their column density distribution determine mean optical depth. Therefore, knowledge of $P(\Delta,z)$ allows to constrain a different combination of the $\Gamma$ and $T_0$, than those inferred from the individual line profiles, i.e. Equations~(\ref{eq:rho-N})--(\ref{eq:tau_Gunn-Peterson}). 

One of the possible ways to infer the gas probability density distribution function is based on the hydrodynamical simulations of the Ly$\alpha$ forest.
Here we used {the analytical} model function from \citet{Miralda-Escude2000,Faucher2008}
\begin{equation}\label{eq:gas_pdf}
    P(\Delta,\,z) = A \exp \left(-\frac{\left(\Delta^{-2/3} -C_0\right)^2}{2(2\delta_0/3)^2}\right) \Delta^{-\zeta},
\end{equation}
{which reasonably well describes the results of simulations.} 
{Fit parameters $A$, $C_0$, $\zeta$, and $\delta_0$}  
{in Equation~(\ref{eq:gas_pdf})} and values of the  effective optical depth  are given in \citep{Faucher2008} for a few redshift points and we interpolated between them.

Thus, a combination of Equations~(\ref{eq:rho-N})--(\ref{eq:gas_pdf}) with measurements of the mean transmission of the Ly$\alpha$ forest makes it possible to estimate the IGM {TDR} parameters  $T_0$ and $\gamma$ and the hydrogen photoionization rate $\Gamma$.

\begin{figure}[t!]
\centering
\includegraphics[width=0.5\textwidth]{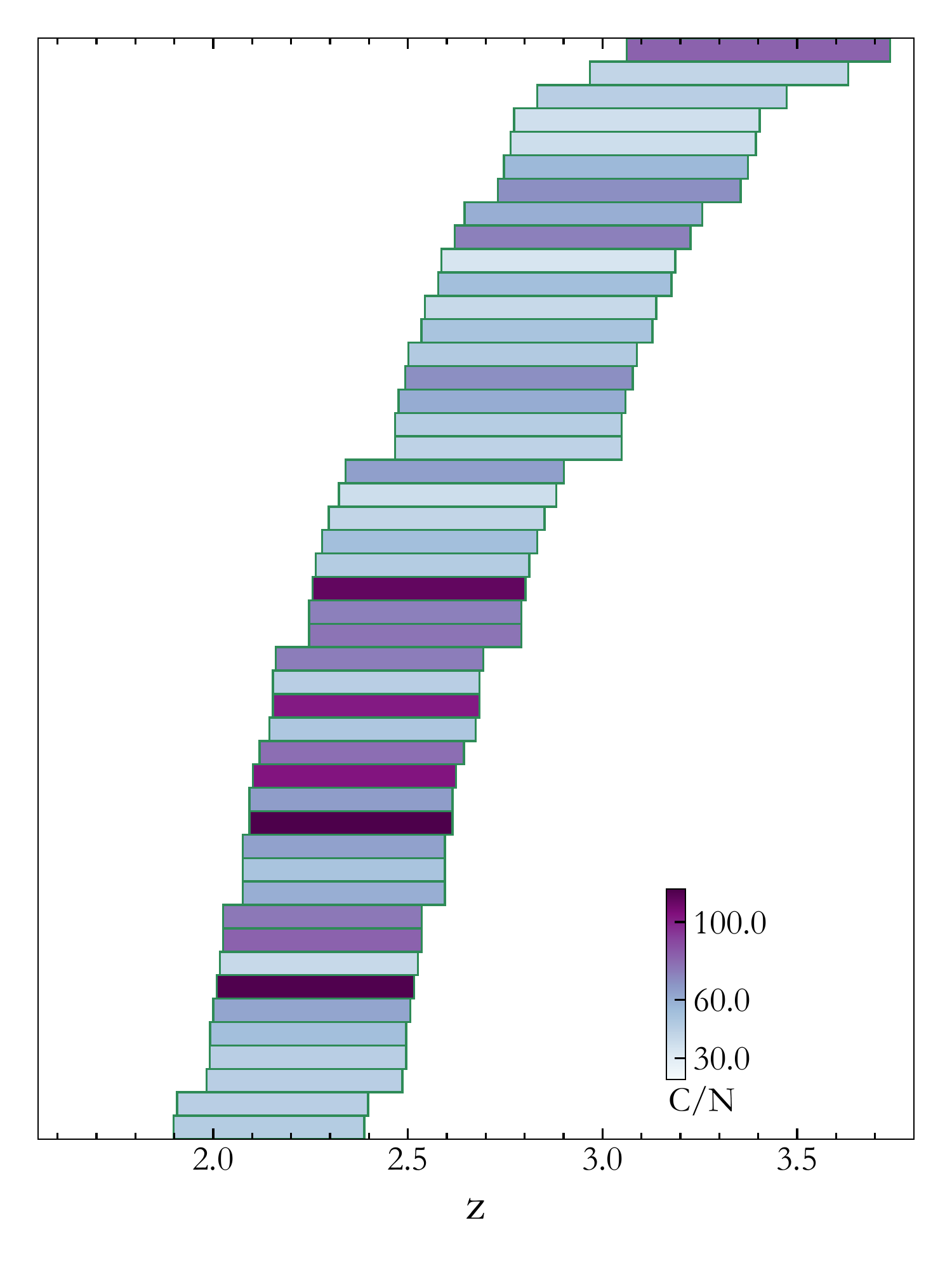}
\caption{Redshift ranges of the spectra used in the analysis. By colors we show the median continuum-to-noise ratio, $C/N$, in 
each spectrum within a given range.
} 
\label{fig:qsos}
\end{figure}

\section{Data}\label{sec:data}
We analysed 47 high-resolution ($R\sim36000-72000$) quasar spectra with high continuum-to-noise ratios ($C/N\sim20-100$) from KODIAQ\footnote{Keck Observatory Database of Ionized Absorption toward Quasars} \citep{OMeara2017}. The redsift range, which corresponds to the Ly$\alpha$ forest in the quasar spectra used, is $z\sim2-4$. All spectra were obtained with the high resolution echelle spectrometer HIRES on the Keck-I telescope. The complete description of the spectra and their parameters are presented in Table~\ref{tab:qsos} in Appendix~\ref{sec:app_data}. Redshift ranges and continuum-to-noise ratios of the quasar sample are illustrated in Figure~\ref{fig:qsos}.

The KODIAQ spectra are already normalized by an unabsorbed quasar continuum. Since it is not straightforward to estimate the quasar continuum in an echelle spectrum, which is   not flux-calibrated usually, we slightly adjusted predefined continuum in each spectrum using a B-spline interpolation over a region free from evident absorption lines. 
Additionally, we renormalized the wings of Ly$\alpha$ lines of the Damped Ly$\alpha$ (DLA) systems\footnote{Saturated Ly-series absorption line systems which have $N \gtrsim 10^{20}$~cm$^{-2}$ and therefore their Ly$\alpha$ absorption lines have prominent smooth wings, which can span up to few tens of \AA.}, that allowed us to include into the analysis the absorption lines located in these regions.
Nevertheless, for the sample analysed in this paper, possible inaccuracies in the continuum determination have only a minor effect on the inferred line parameters.

\section{Quasar spectra analysis}\label{sec:fitting procedure}

\begin{figure*}[htb]
\centering
\includegraphics[width=1.0\textwidth]{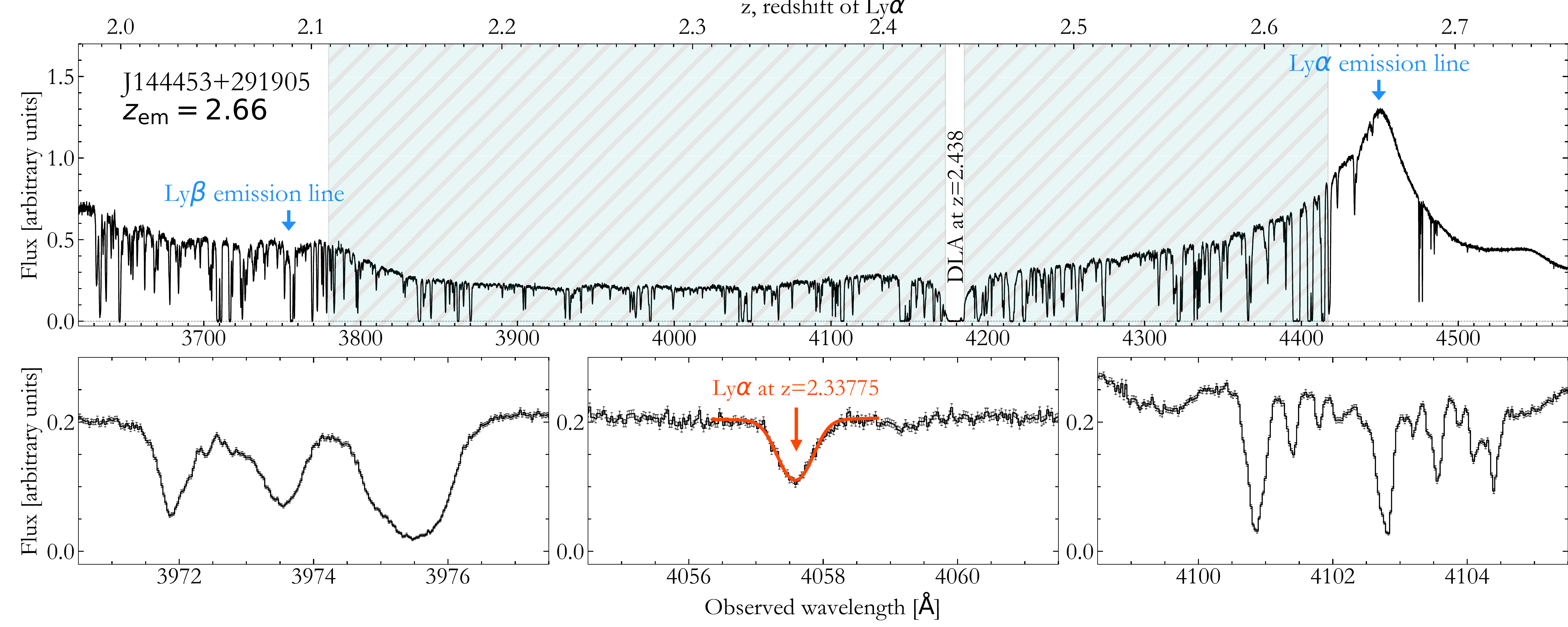}
\caption{Spectrum of the quasar J144453+291905  obtained with HIRES on the Keck telescope. The spectrum is shown by the black lines in the panels. The spectrum is not flux-normalized. {\sl Top panel}: the part of the spectrum containing Ly$\alpha$ forest. The hatched region indicates the wavelength range (i.e. redshift range, see labels at the top axis) located between quasar Ly$\alpha$ and Ly$\beta$ emission lines used for Ly$\alpha$ forest analysis.  {\sl Left bottom panel}: an example of a complex Ly$\alpha$ forest line consisting of several individual absorption lines blended with each other. {\sl Central bottom panel}: an example of a simple ``solitary-like'' Ly$\alpha$ forest line used in our study. The red line shows the Voigt profile fit to the Ly$\alpha$ absorption line at $z\approx 2.33775$. {\sl Right bottom panel}: an example of a metal lines profile. These lines are \SiII$\lambda1193$\AA\ absorption lines from the DLA system at $z\approx2.4366$ (damped Ly$\alpha$ absorption line can be seen in the top panel at $\sim4180$\AA).
} 
\label{forest}
\end{figure*}

{
The traditional approach for the analysis of the Ly$\alpha$ forest lines in quasars spectra is the multicomponent fitting with, e.g., the \textsf{VPFIT} software \citep{Carswell2014}. 
In this approach, individual components are assumed to be formed in spatially separated regions and each of them is described by the Voigt profile (with different $z$, $N$, and $b$). An example of complex Ly$\alpha$ line is shown in the left bottom panel of Figure~\ref{forest}.
However, the decomposition of such absorption profiles is a non-trivial inverse problem which can lead to ambiguous and biased results. 
To avoid these complications, we focus on the selection of the Ly$\alpha$ lines with the simple structure dominated by a single component.
Moreover, the analytical approach described in the previous Section~\ref{sec:igm_eos}, by construction, directly applies to the separate IGM filaments,
unaffected by a presence of adjacent filaments.\footnote{
The presence of adjacent filaments may indicate the regions with rich cosmological structure, and, as a consequence, enhanced density of ionizing sources, resulting in higher UV background than the mean one).
} Therefore the choice of solitary lines for the analysis in our formalism is natural to avoid possible bias.}

We developed a custom routine to select and determine parameters of the Ly$\alpha$ forest lines with simple ``solitary-like'' structure as detailed below \citep{Telikova2018a,Telikova2018b}. In each quasar spectrum with the redshift $z_{\rm QSO}$, we selected the wavelength region between Ly$\alpha$ and Ly$\beta$ emission lines to search for the Ly$\alpha$ forest lines. We made a blueward offset from the Ly$\alpha$ emission line with the velocity magnitude $v_{\rm red}=3000$\,km\,s$^{-1}$, i.e. $1+z_{\rm red} = (1 + z_{\rm QSO}) (1 - v_{\rm red}/c) $.
This offset is required to exclude possible broad absorption line systems associated with the quasar and remove Ly$\alpha$ forest lines associated with the quasar proximity zone. In a similar way, the position of the blue boundary of the selected wavelength region is given by $\lambda_{\rm blue}=\lambda_{\rm Ly\beta} (1+z_{\rm blue})$, where $\lambda_{\rm Ly\beta}=1025.72$\,\AA\ is the wavelength of the Ly$\beta$ line in the rest frame and $z_{\rm blue}$ is determined using the redward offset $v_{\rm blue}=-2000$\,km~s$^{-1}$. This guarantees  exclusion of the Ly$\beta$ lines from potential absorption systems with negative peculiar velocities in the quasar rest frame. The resulting search region should contain only  Ly$\alpha$ absorption lines 
with some contamination from metal lines (see Section~\ref{metals}), and does not contain any lines of the higher Lyman series. An example of the spectrum and corresponding search region are shown in the top panel of Figure~\ref{forest}. We also masked evidently bad regions and the regions near  Ly$\alpha$ absorption lines from DLAs. 

\begin{figure*}[ht!]
\centering
\includegraphics[width=1.0\textwidth]{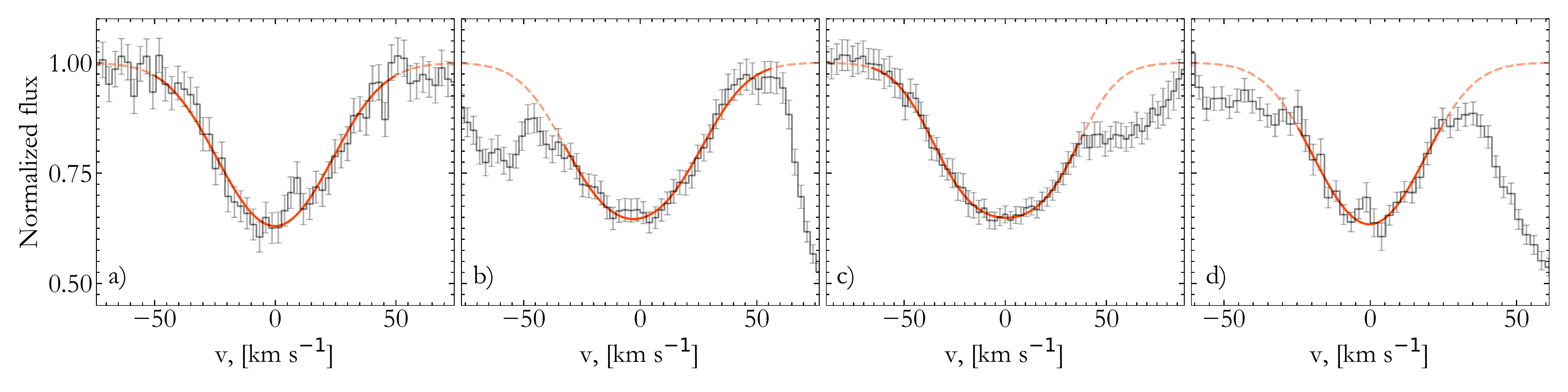}
\caption{Types of masks used in the spectral analysis. The dashed and solid red lines show the total and masked fit profiles, respectively, overlaid on the examples of partially blended Ly$\alpha$ lines from an actual spectrum (gray error bars). Panels a)--d) correspond to four lines in Equation~(\ref{eq:mask}).} 
\label{fig:mask}
\end{figure*}

We limited our analysis of the Ly$\alpha$ lines to a box in $(N,\ b)$ parameter space, which contains the expected location of the cutoff of the distribution for selected redshift ranges. This box was specified as\footnote{In fact, to avoid the edge effects, the upper limit for b was extended to $b=50$ km~s$^{-1}$. After searching and fitting procedure, we picked lines  with $b< 30$ km~s$^{-1}$ for further analysis.}
\begin{equation}\label{eq:box}
\begin{cases}
	\log N\, [\mbox{cm}^{-2}] = 13 - 14.5, \\
    b = 10 - 30\,\mbox{km~s$^{-1}$}.
\end{cases}
\end{equation}
The lower limit for $b$ was chosen as $10$~km~s$^{-1}$, since at lower $b$ the majority of the lines originate from the transitions in various metals instead of the  Ly$\alpha$ forest. 
Upper limit for $b$ and lower limit for $N $ were chosen to avoid problems with identification of broad or/and weak lines. Upper limit for $\log N [\mathrm{cm}^{-2}]$ is placed at $14.5$ since the number of lines with larger $N$ is small (due to a power-law distribution of the absorbing systems over $N$) and also since an analysis of the highly saturated lines (with $\tau_0 \gg 1$) is complicated. For each spectrum, we generated synthetic Ly$\alpha$ lines on a dense grid of $N$ and $b$, where grid steps were estimated  taking into account the resolution and signal-to-noise ratio in the spectrum based on the Fisher matrix calculations (see Appendix~\ref{sect:fisher} for details). 
The resulting synthetic line profiles were matched (as detailed below) with the observed spectra within the search redshift region using the redshift step size corresponding
to the typical pixel size. This allowed to minimize the computational cost during the scanning procedure. Since Ly$\alpha$ lines are much wider 
than the instrument response function 
there are at least several pixels within the profile of each line. This makes missing any of the lines via the described procedure highly unlikely.

For each point $(z,\, N,\, b)$ in the parameter space we calculated the likelihood  ${\cal L}$ to match the spectrum with the model profile assuming the Gaussian distribution of spectral uncertainties\footnote{This does not take into account that the adjacent pixels in the spectrum can be correlated.}
\begin{equation}
	\label{eq:likelihood}
2 \ln{\mathcal L} \equiv -\chi^2 = -\sum\limits_{i=1}^{n}\left(\frac{y_i - I_{\rm conv}(\lambda_i)}{\sigma_i}\right)^2,
\end{equation}
where $y_i$ is a flux in spectral pixels specified by $\lambda_i$ with measurement uncertainties $\sigma_i$. The sum is carried over all pixels selected to compare the observed spectrum with the absorption line model profile.

At each position $(z,\, N,\, b)$  in the parameter grid, we selected the pixels based on the model profile using 4 types of mask:
\begin{widetext}
\begin{equation}
\label{eq:mask}
  M =
    \begin{cases}
      \tau(\lambda) > 0.05,  \\
      \tau(\lambda) > 0.05,\,\, \lambda > \lambda_\mathrm{Ly\alpha} (1+z)\,\, \mbox{and}\,\, e^{-\tau(\lambda)} > e^{-\tau_0} / 2,\,\, \lambda < \lambda_\mathrm{Ly\alpha} (1+z),   \\
      \tau(\lambda) > 0.05,\,\,\lambda < \lambda_\mathrm{Ly\alpha} (1+z) \,\,\mbox{and}\,\, e^{-\tau(\lambda)} > e^{-\tau_0} / 2,\,\, \lambda > \lambda_\mathrm{Ly\alpha} (1+z),   \\
      
      e^{-\tau(\lambda)} > e^{-\tau_0} / 3,
    \end{cases}       
\end{equation}
\end{widetext}
where $\tau(\lambda)$ corresponds to the model profile at a certain $(z,\, N,\, b)$ point, $\tau_0$ is the optical depth at the line center, see Equation~\eqref{eq:tau}.
These masks allow to include into analysis the lines which are partially blended in the blue or red wings (the second and the third masks in Equation~\eqref{eq:mask}, respectively) and simultaneously in both wings (the forth mask in Equation~\eqref{eq:mask}). A solitary line without blends can be well fitted with the first mask in Equation~(\ref{eq:mask}). 
These four masks and examples of fitted lines are shown in Figure~\ref{fig:mask}.

For each mask, we find the local maxima of the likelihood function on the calculated grid in the parameter space which satisfy a criterion 
$\chi^2_\mathrm{red}\equiv \chi^2 / \mathrm{dof} < \chi^2_{\rm thres}$, where $\mathrm{dof} = n - 3$ and $n$ is the number of pixels in the masked line profile.  
Using this criterion, we select the lines that can be well fitted with the Voigt profile. 
We empirically found running a few spectra that the threshold value of 
$\chi^2_{\rm thres}=3$ 
provides a good trade-off between the completeness of the sample (regarding the selection of lines with a simple structure) and the fraction of poorly fitted lines. We additionally checked that there is no degeneracy between  $b$ and $N$ parameters in our sample. 
On the other hand, a certain line in the spectrum can be found during the search with more than one mask. The evident example is the line which is found with the whole mask (first mask in Equation~\eqref{eq:mask}). Indeed, it most probably will be found with three other masks as well. In such case, we kept only one instance in the sample using the following order of preference: the whole mask a), the left wing b) or the right wing c), and the central part d),  see Figure~\ref{fig:mask}.

\begin{figure}[t]
\centering
\includegraphics [width=0.5\textwidth]{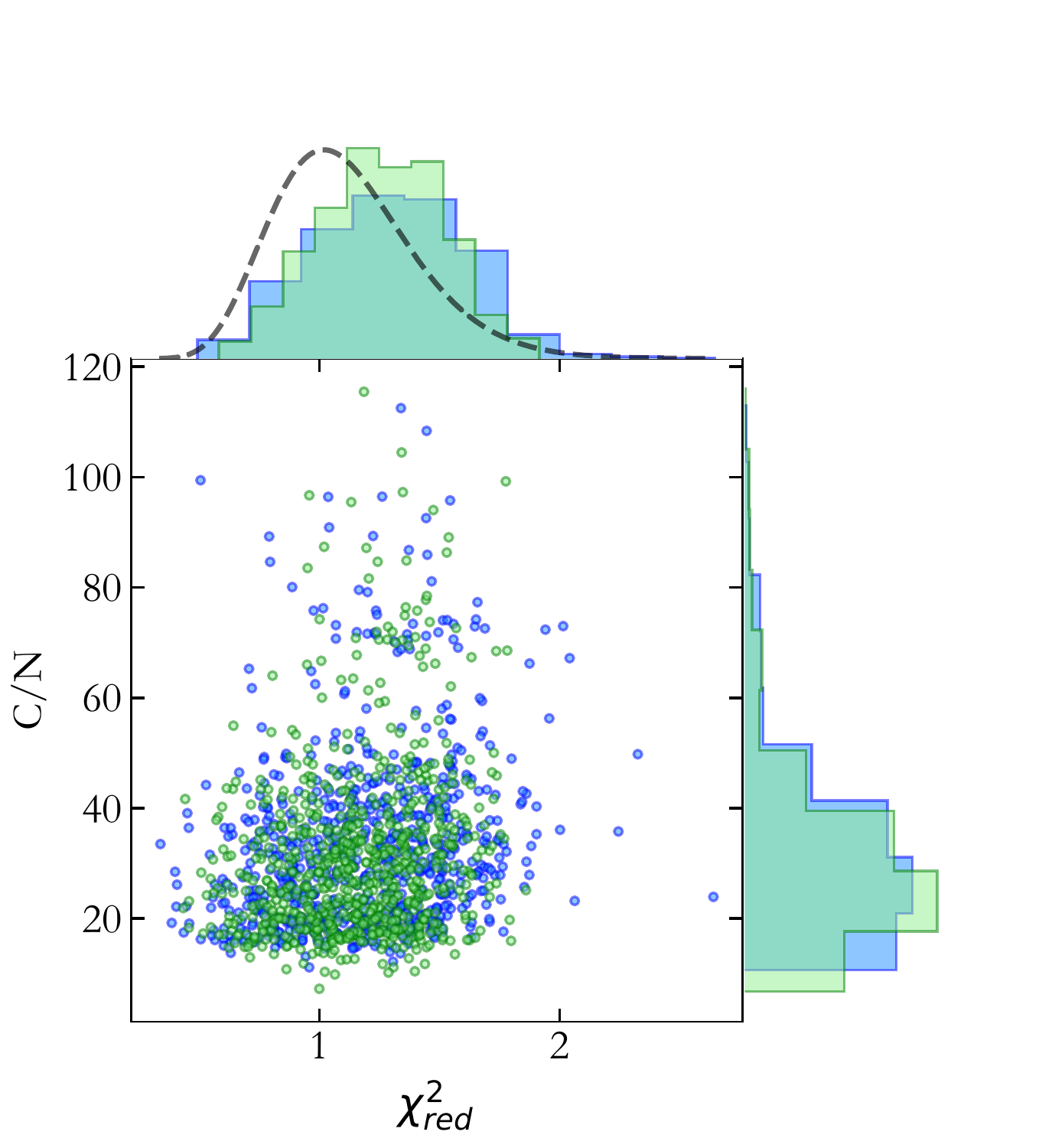}
\caption{Continuum to noise ratio {\it vs.} $\chi_{\rm red}^2$ for the {full} metal-cleaned sample.
The green and blue points correspond to the lines which were fitted with the whole mask and others mask, respectively. The dashed line corresponds to the probability density function of the $\chi^2$ distribution with 20 degrees of freedom. 
}
\label{fig:sample_stats}
\end{figure}

After locating all local maxima, for each of them we run a likelihood maximization with the selected mask. The uncertainties of the line parameters were estimated as the square roots of the diagonal elements of the covariance matrix calculated for best-fit parameters. 
For each potential Ly$\alpha$ line found in this way, we also checked that the corresponding higher Lyman series lines (e.g., Ly$\beta$) do not contradict the spectrum.
At a final step, we selected for further analysis the lines which have  best-fit $\chi^2_\mathrm{red}$ value within the 99\% confidence interval of the $\chi^2$ distribution with respective number of dof.
As a result, for each quasar we obtained the sample of the Ly$\alpha$ forest lines which have simple structures of the line profiles. After the  metal lines rejection procedure, which is discussed in Section~\ref{metals}, the final sample contains 1503 Ly$\alpha$ lines, among which 786, 467, and 271 lines correspond to the masks a), b) or c), and d), respectively (see Figure~\ref{fig:mask}).  The use of the different masks allowed us to include partially blended lines which increased the sample by a factor of two. Distribution of the lines over $\chi^2_\mathrm{red}$ and continuum to noise ratio, $C/N$, are shown in Figure~\ref{fig:sample_stats}, where green and blue colors correspond to the lines fitted with the whole mask, a), and the other masks, respectively.  
The distribution of $\chi^2_{\rm red}$ for the obtained lines is in a good agreement with the  $\chi^2$ distribution for 20 degrees of freedom\footnote{Notice, that the actual number of the degrees of freedom for lines in the sample vary between $\sim15-50$.} (dashed curve in Figure~\ref{fig:sample_stats}). This agreement supports the adopted choice of the line selection criteria.  

\section{$N-b$ distribution analysis}\label{sec:analysis of the b-N sample}

\subsection{Metal lines rejection}
\label{metals}

\begin{figure*}[ht]
\centering
\includegraphics [width=1.0\textwidth]{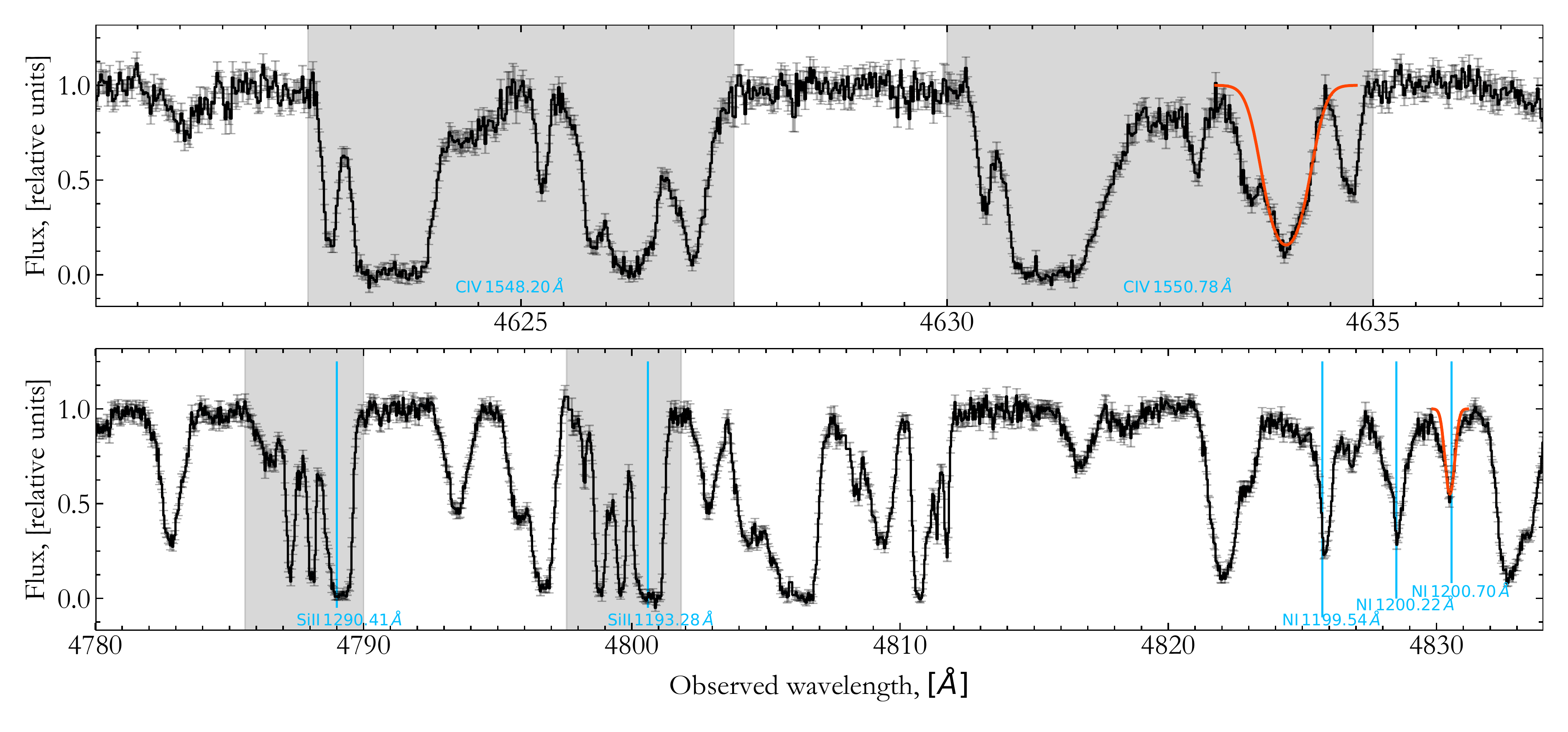}
\caption{{\sl Top panel:} The portion of a quasar spectrum containing \CIV\ doublet at $z\approx 1.987$. It is clearly seen that doublet lines have the same velocity structure (extended over $\sim 500$~km~s$^{-1}$, as shown by the filled regions), 
which makes possible its reliable identification.  {\sl Bottom panel:} The portion of a quasar spectrum containing metal lines associated with a DLA system at $z\approx 3.0229$. In both panels, false-positively identified Ly$\alpha$ lines are shown by the red curves.}
\label{fig:doublets}
\end{figure*}

It is well known, that there are various metal absorption lines (that have restframe wavelengths in UV and optical bands) in the quasar spectra. Such metal absorption lines can be easily identified in the spectral region redwards the Ly$\alpha$ emission line, that is free from the Ly$\alpha$ forest lines. Unfortunately, they are present inside the Ly$\alpha$ forest wavelength range and contaminate our sample \citep[e.g.][]{Rudie2013,Hiss2017}. Therefore, it is necessary to clean the sample from them.

The metal absorption lines usually demonstrate a complex structure, i.e. several components can be clearly distinguished in their profiles. The dominant majority of metal absorption lines in quasars spectra are doublets, such as {\CIV}$\lambda\lambda$1550,1548\AA,  {\SiIV}$\lambda\lambda$1393,1402\AA, {\MgII}$\lambda\lambda$2803,2796\AA, and metal lines associated with DLA (Damped Ly$\alpha$) systems, sub-DLA (sub-Damped Ly$\alpha$), and Lyman Limit Systems. Usually a given metal absorption line in the spectrum has at least one ``counterpart'', the metal line with a similar structure but at a different wavelength. In principle, this allows one identify such lines in the spectrum. Unfortunately, the wavelength separation in doublet metal lines is usually small, therefore the ``counterparts'' fall in the region of Ly$\alpha$ forest and can be by a chance blended with Ly$\alpha$ forest lines. 
The metal lines connected to the DLA system usually show similar velocity structures in groups of low-ionized species (such as {\SiII}, {\OI}, etc.)  or in groups of high-ionized species (such as {\SiIV}, {\CIV}), allowing their identification. However, there exist ``solitary'' metal lines, such as {\SiIII}$\lambda$1206\AA, which, although being connected to the DLA, have no counterparts with similar velocity structure. 
All these factors preclude a straightforward automatization of the metal lines rejection procedure.

\begin{figure*}[ht]
\centering
\includegraphics [width=\textwidth]{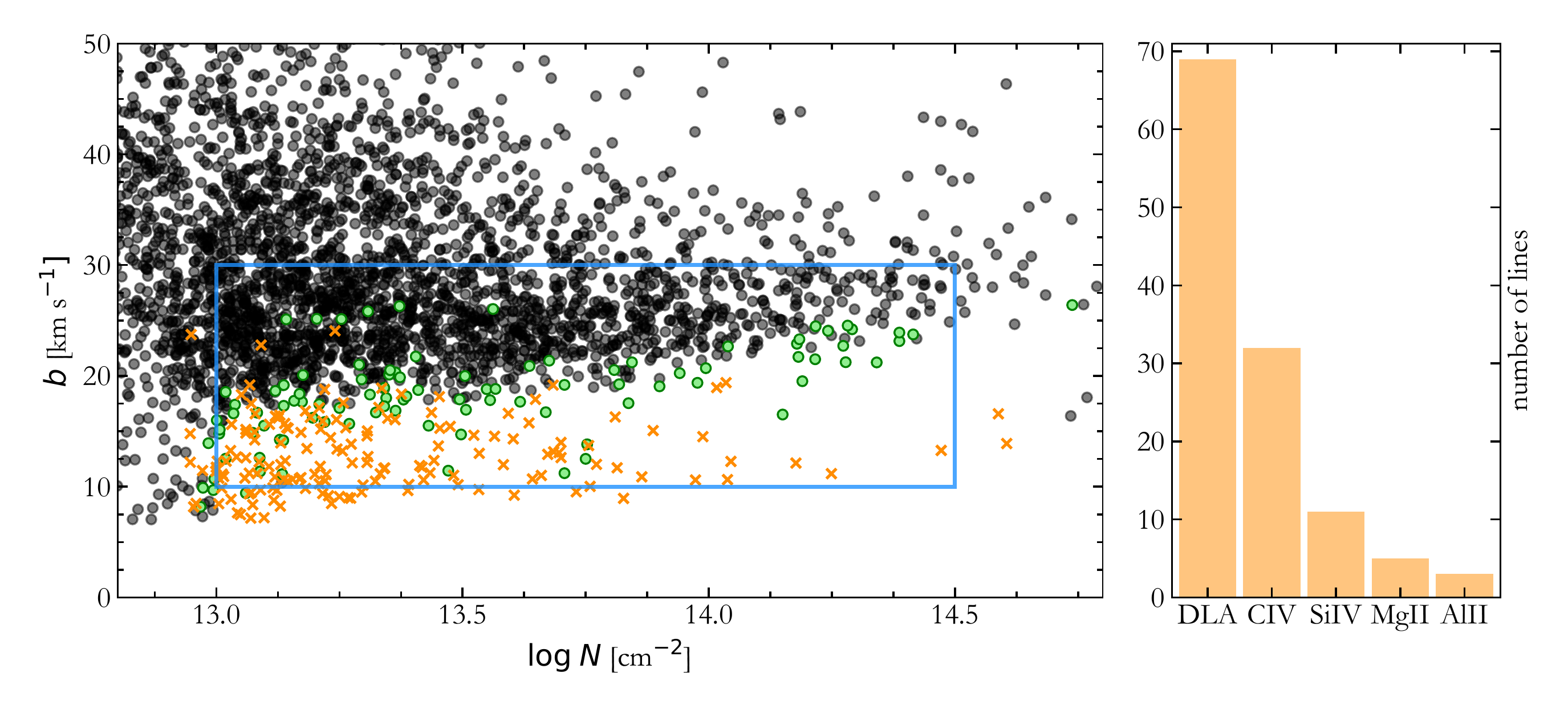}
\caption{{\sl Left panel:} The distribution of $(b, N)$ parameters for the {full sample of} lines selected by our searching procedure.
The orange crosses show the identified metal lines. The green points indicate the Ly$\alpha$ lines, which parameters gives agreed line profiles in higher Lyman series lines. Notice, that we randomly checked these lines for this only in the vicinity of lower cutoff of $(b, N)$ distribution. The blue rectangle indicates the box used for the fit of the $(b, N)$ distribution. {\sl Right panel:} number of identified metal lines binned by their origin.
}
\label{fig:sample_rej}
\end{figure*}

Therefore, we visually inspected the sample looking for doublet metal lines (associated or not with DLAs) based on the similarities in the velocity structure of the doublet components (see Figure~\ref{fig:doublets}) with the expected wavelength separation taken from the metal lines database. We found 120 metal lines in the initial sample near the low boundary of $b-N$ distribution.
These lines are shown by orange crosses in the left panel of Figure~\ref{fig:sample_rej}. 
The redshift range of the final Ly$\alpha$ lines sample is $z=1.9-3.7$.

\subsection{Maximum Likelihood Estimation}

We developed a new statistical method for analysing the full $N-b$ sample \citep{Telikova2018a,Telikova2018b}. This method is different from the most of previously used methods that mainly focus only on the low boundary position \citep{Schaye1999,Bolton2014,Hiss2017}, see, however, \citet{Ricotti2000,Hiss2019}.

The total broadening of a line contains contributions from the thermal broadening and other additional broadenings. Neglecting broadening due to cosmological expansion (which impacts only the absorbers with lowest column densities \citealt{Garzilli2015, Garzilli2018}), the additional broadening is mostly due to peculiar motions inside an absorber, emergent from either turbulent motions or projection effects. Such broadening is usually assumed to be Gaussian 
with some broadening parameter $b_{\rm add}$. 
We further assumed that thermal and peculiar motions are uncorrelated, then 
the total line broadening has a form 
\begin{equation}\label{eq:b_total}
 b^2  = b^2_{\rm th} + b^2_{\rm add},
\end{equation}

Assuming that $N$ and $b_\mathrm{add}$ are independent random variables, the
2D joint probability distribution function of Ly$\alpha$ absorbers over $N$ and $b$ can be written as
\begin{equation}\label{eq:pdf_N_b}
  f(N,\, b)=\int f_N(N)f_{\rm add}(b_{\rm{add}})\delta \left(b-\sqrt{b_{\rm{th}}^2+b_{\rm{add}}^2} \right)\rm{d} \emph{b}_{\rm{add}},
\end{equation}
where $f_N(N)$ is a probability density distribution over the column density and $f_{\rm add}(b_{\rm{add}})$ is a probability density distribution over the additional broadening parameter.

\begin{figure*}[th]
\centering
 \includegraphics [width=1.0\textwidth]{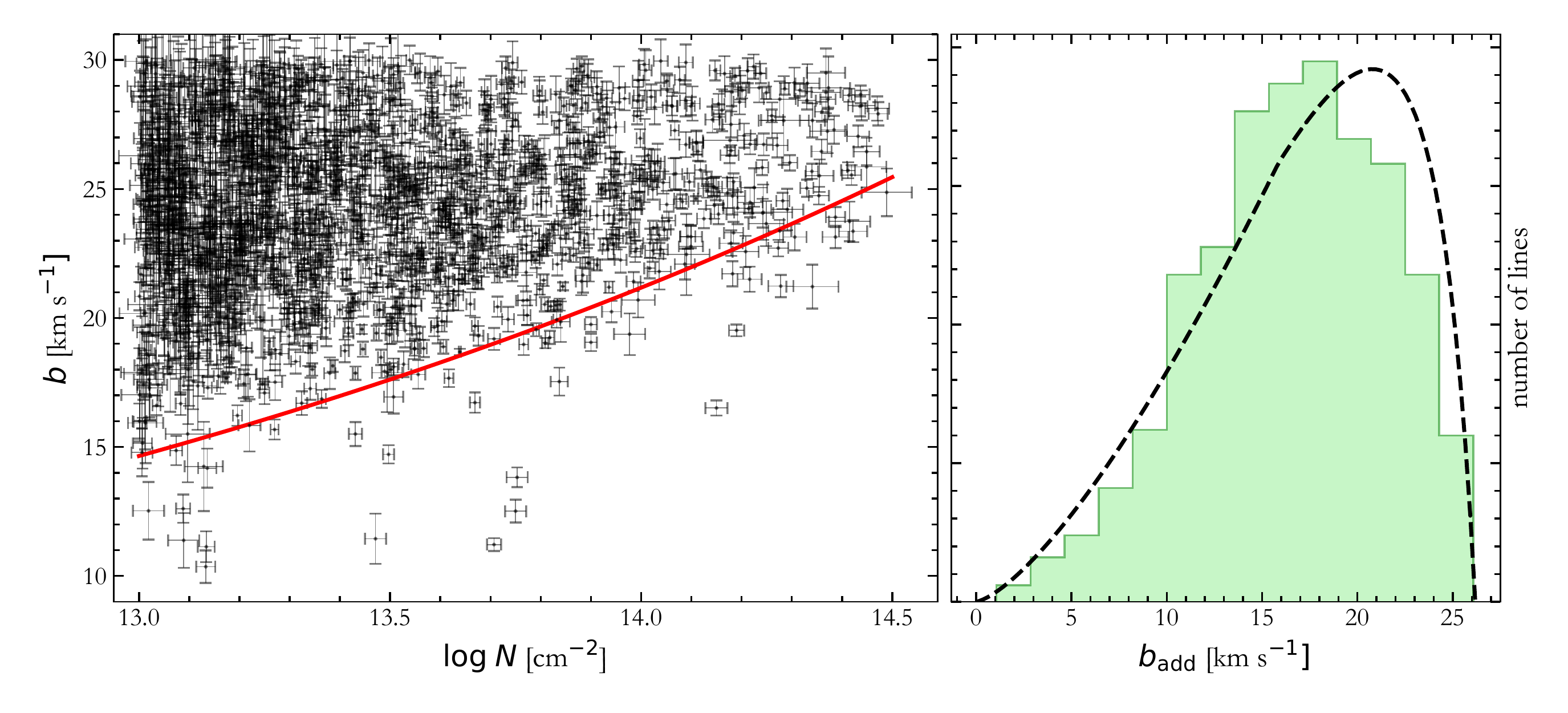}
\caption{{\it Left panel:} Ly$\alpha$ forest lines after metal lines rejection 
{(full sample)} are shown by black error crosses. Red line is the low boundary of this sample plotted by eye.\\ {\it Right panel:} {Estimated distribution of the additional broadening 
$b_\mathrm{add}$ for this sample, assuming average redshift $\bar  z = 2.67$ and the by-eye thermal contribution (red line). The points below the red line are disregarded, see text for details. }
Black dashed curve is $b_{\rm add}$ distribution function calculated for  $p=1.4$ and $\beta=-1.4$.}
\label{fig:b_turb_hist}
\end{figure*}

The column density distribution is well-studied in a wide redshift range and has a power-law form $f_N(N) \propto N^{\beta}$ \citep[e.g.][]{Janknecht2006J,Rudie2013,Kim2002}. The additional broadening distribution, $f_{\rm add}(b_{\rm{add}})$, is unknown, and we assume that it has a power-law shape $\propto b_{\rm add}^{p}$. We based this choice on the following empirical arguments.
 Drawing the lower envelope in the $N-b$ plane by hand (red solid line in the left panel of Figure~\ref{fig:b_turb_hist}) and ignoring lines which fall below this putative envelope, we can obtain approximate ``empirical'' values of $b_\mathrm{add}$ for each line. The resulting distribution of $b_\mathrm{add}$ is shown in the right panel of Figure~\ref{fig:b_turb_hist} \footnote{We employed the average redshift $\bar z = 2.67$, however, the shape of the distribution does not change significantly if individual redshifts of the absorbers are taken into account.}. It is necessary to stress, that this distribution does not reflect completely the distribution $f_\mathrm{add}$, because large values of $b_\mathrm{add}$ are undersampled in the right panel of Figure~\ref{fig:b_turb_hist} due to the truncation provided by the selection box, given in Equation~(\ref{eq:box}). Only at low values of $b_\mathrm{add}\lesssim 15$~km~s$^{-1}$, the empirical distribution over $b_\mathrm{add}$ coincides with $f_\mathrm{add}$, while its bell-like shape at large $b_\mathrm{add}$ is due to selection effects. 
 
Indeed, the empirical distribution $f_\mathrm{em}(b_\mathrm{add})$ can be written as the marginalization of the 2D distribution in $N-b_\mathrm{add}$ over $N$ in the box, i.e., omitting the normalization constant, 
\begin{equation}
    f_\mathrm{em}(b_{\rm add}) \propto f_\mathrm{add}(b_\mathrm{add}) \int_{N_{\rm low}}^{N_{\rm up}} f_N(N) {\rm d}N,
    \label{eq:f_add_marg}
\end{equation}
where $\log N_{\rm low}[\mathrm{cm}^{-2}] = 13$ (see Equation~(\ref{eq:box})), and the upper limit  $\log N_{\rm up}=\min(14.5,\ \log N_\mathrm{max}(b_\mathrm{add}))$. Here, $N_\mathrm{max}(b_\mathrm{add})$ is a solution of the equation
\begin{equation}\label{eq:Ntrunc}
    b_{\rm th}(N_{\rm max}) = \sqrt{b_{\rm max} - b_{\rm add}},
\end{equation} 
where $b_{\rm max} =30$\,km\,s$^{-1}$ is the upper boundary of our sample, Equation~(\ref{eq:box}). 
Therefore, $f_\mathrm{em}$ is equal to $f_\mathrm{add}$ up to values of $b_\mathrm{add}$ corresponding to $\log N_\mathrm{max} [\mathrm{cm}^{-2}]=14.5$ in Equation~(\ref{eq:Ntrunc}). For these $b_\mathrm{add}\lesssim 15$~km~s$^{-1}$, the empirical distribution in the right panel of Figure~\ref{fig:b_turb_hist} has a power-law shape justifying our anzatz. 
For illustration, we show $f_{\rm em}(b_\mathrm{add})$ calculated using Equation~\eqref{eq:f_add_marg} with $p=1.4$ and $\beta=-1.4$ by the dashed black line in the right panel of Figure~\ref{fig:b_turb_hist}. The plot indicates a good agreement of the proposed distibution shape with the data. {Several illustrative examples of the dependence of the adopted distribution shape on the model parameters are shown in Appendix~\ref{sect:pdf_var_parameters}.}

The likelihood for observing the given datapoint $(N_i,\,b_i)$ from the distribution \eqref{eq:pdf_N_b} should account for the measurement errors and selection effects arising from the restrictions specified by the box, Equation~\eqref{eq:box}. The appropriate likelihood function is
\begin{widetext}
\begin{equation}\label{eq:L_data}
{\cal L_{\rm data}}(N_i,\, b_i) =  \frac{\int f(\widetilde{N},\, \widetilde{b}) \exp(-{(\widetilde{N}-N_i)^2}/{2\sigma_{N_i}^2}) \exp(-{(\widetilde{b}-b_i)^2}/{2\sigma_{b_i}^2}) \,{\rm d}\widetilde{N}\,{\rm d}\ \widetilde{b}}
{\int f(\widetilde{N},\, \widetilde{b})I(N, b) \exp(-{(\widetilde{N}-N)^2}/{2\sigma_{N_i}^2}) \exp(-{(\widetilde{b}-b)^2}/{2\sigma_{b_i}^2}) \,{\rm d}N\,{\rm d}b\,{\rm d}\widetilde{N}\,{\rm d}\widetilde{b}},
\end{equation}
\end{widetext}
where $\sigma_{N_i}$ and $\sigma_{b_i}$ are the measurement uncertainties obtained from the fitting procedure, see Section~\ref{sec:fitting procedure}, and the function $I(N,\,b)$ in the normalizing denominator is an indicator function corresponding to Equation~\eqref{eq:box}. For the numerator, the indicator function $I(N_i,\,b_i)$ is always unity by definition and is omitted. The proper normalization of the likelihood is important when the truncated sample problems are considered  \citep[e.g.,][]{Hogg2010}.

In the left panel of Figure~\ref{fig:b_turb_hist} there exist several ``outlier'' points, i.e. the lines falling much below the approximate lower boundary of the distribution even if measurement uncertainties and redshift corrections (see Equation~(\ref{eq:b-N})) are taken into account. These outliers can be unidentified metal lines 
or the Ly$\alpha$ systems with peculiar properties (i.e. much lower temperature than follows from the general law in Equation~(\ref{eq:eos})). 
These points can not be properly taken into account by the likelihood (\ref{eq:L_data}). {Here, 
following
our previous work  \citep{Telikova2018b},} we use the mixture model, where each point can be generated either from the data distribution with the likelihood given by Equation~\eqref{eq:L_data} or from the outlier distribution with some likelihood function ${\cal L}_\mathrm{out}$ \citep[see][for the details]{Hogg2010}.
In this approach, the total likelihood is
\begin{equation}\label{eq:L}
    {\cal L} =\prod\limits_{i} \left[ (1-P_b) {\cal L_{\rm data}}(N_i,\, b_i) + P_b{\cal L_{\rm out}}(N_i,\, b_i) \right],
\end{equation}
where 
$P_b$ is a probability of any random point from the sample to be an outlier.
 In general, the distribution of outliers is unknown and one needs to assume some reasonable distribution that is capable to catch the longer outlier ``tails'' missed by the model distribution. Following \citet{Hogg2010}, we first tried the Gaussian distribution for outliers, but the posterior distributions for mean and variance demonstrated that the resulting distribution was close to uniform. Therefore, for the final fit we chose a uniform distribution for outliers to reduce the number of fit parameters and improve convergence.

\subsection{MCMC}\label{subsect: MCMC}

\begin{figure*}[ht]
\centering
\includegraphics [width=0.95\textwidth]{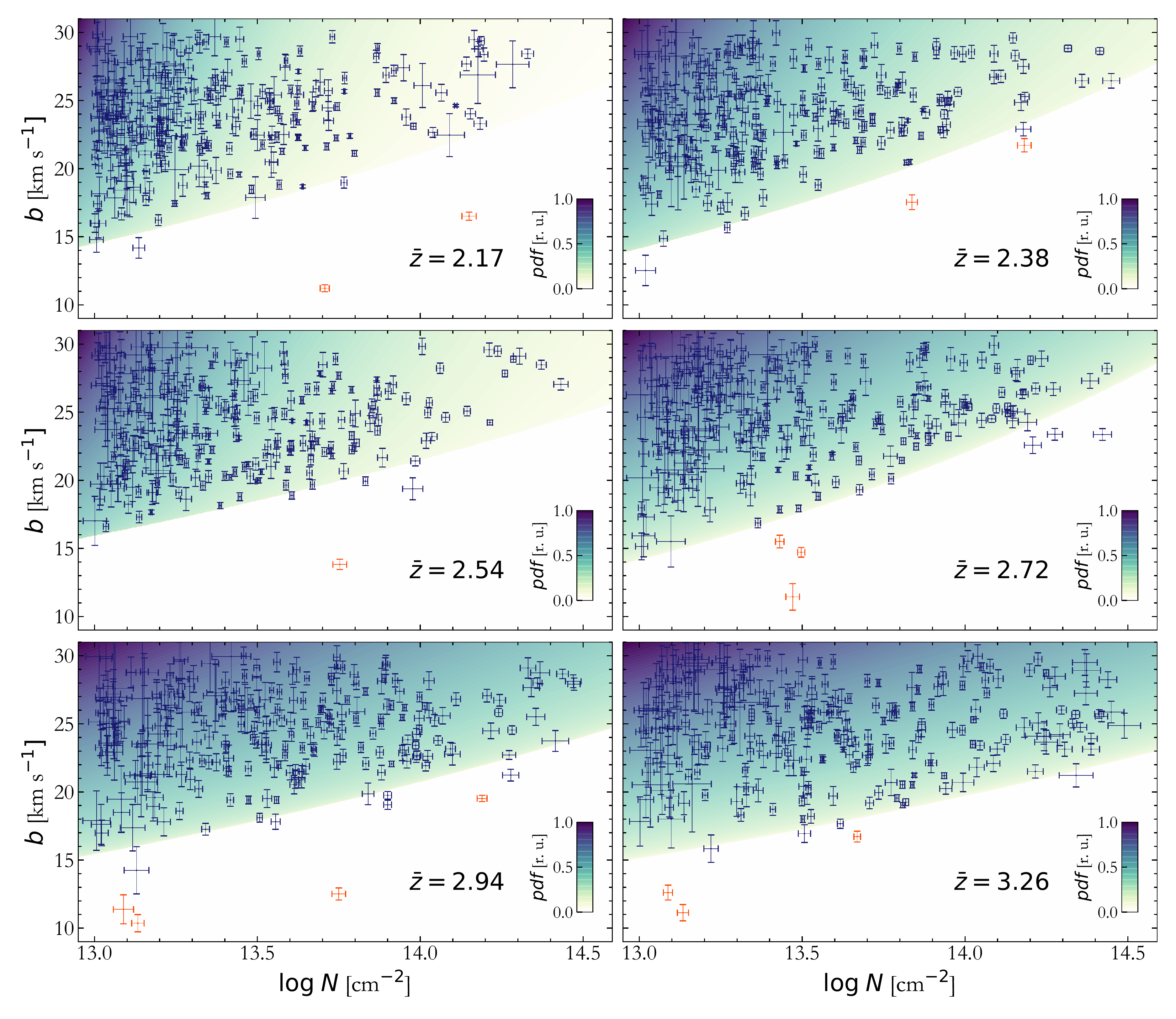}
\caption{Blue error crosses show the Ly$\alpha$ lines sample after metal lines rejection. Best-fit of the distribution function in each redshift bin is shown by the colour gradient area. Red error crosses are outliers.}
\label{fig:data_pdf}
\end{figure*}
We split our Ly$\alpha$ lines sample into 6 redshift bins, which contain approximately equal number of lines and analyzed each bin individually.
The likelihood in Equation~\eqref{eq:L} depends on 5 model parameters: four parameters correspond to the Ly$\alpha$ forest absorption systems ($\xi,\, \log b_0,\, \beta,\, p$) and one parameter, $P_b$, describes outliers. Notice, that the thermal broadening $b_\mathrm{th}$ in Equations~(\ref{eq:pdf_N_b})--(\ref{eq:L}) is calculated employing the actual redshift of each line, see Equation~(\ref{eq:b-N}).
We obtained the posterior distribution of the fit parameters in the Bayesian framework using the affine Markov Chain Monte Carlo (MCMC) sampler \textsf{emcee} \citep{Foreman-Mackey2013}. 
For all parameters we chose flat priors.
We set a physically motivated upper boundary $P_b<0.5$, because we suppose that most of the lines in our sample are not outliers.
Posterior distributions of the parameters estimated for individual redshift bins are shown in Appendix~\ref{sect:posteriors} in Figure~\ref{fig:triangles}. Fit results for the parameters of the adopted model are shown in Table~\ref{tab:pars}. The point estimates are medians of the posterior distributions and uncertainties of the parameters correspond to the 68\% credible intervals. 
The model distributions calculated with the parameters from Table~\ref{tab:pars} are overlaid (gradient-filled areas) in Figure~\ref{fig:data_pdf} on the Ly$\alpha$ forest samples (blue crosses) for each bin; the mean redshift in a bin is indicated in the corresponding panel of the figure.
The red crosses in Figure~\ref{fig:data_pdf} indicate  most probable outliers, i.e. the points for which the expectation value for  the first term in  Equation~(\ref{eq:L}), calculated with respect to the posterior distribution, is less than the expectation value for the second term. 
{We also checked the performance of the MCMC procedure by applying it to a number of mock Ly$\alpha$ samples generated from the distribution in Equation~(\ref{eq:L}) (including outliers distribution).
Running this procedure few tens of times, we found that the resulting credible intervals in expected number of cases cover the initial values of the parameters without showing significant outliers. This assures the robustness of the  developed procedure. }

{In our preliminary work \citep{Telikova2018b}, we performed the similar fit to the data, albeit tying the parameters $p$ and $\beta$ (and also the outlier fraction $P_b$) between the redshift bins to reduce the uncertainties of the fit. Further analysis has  reveled, that this was erroneous assumption, since, as the Table~\ref{tab:pars} shows, these parameters, in fact, change significantly from bin to bin. This is especially prominent for the  parameter $\beta$.} 
{Therefore in the following we do not compare the present results with those of \cite{Telikova2018b}}.

\begin{deluxetable*}{*{7}{l}|ll}
\tablecaption{Model parameters.\label{tab:pars}}
\tablehead{
\colhead{redshift range} & \colhead{$\bar{z}$} & \colhead{$\xi-1$} & \colhead{$\gamma-1$} & \colhead{$\log b_0$} & \colhead{$p$} & \colhead{$\beta$} & \colhead{$T_0$} & \colhead{$\Gamma$}  \\
\colhead{} & \colhead{} & \colhead{} & \colhead{} &\colhead{km~s$^{-1}$} & \colhead{} & \colhead{} & \colhead{$10^4$ K} &\colhead{$10^{-12}$ s$^{-1}$} 
}
\startdata
$1.90-2.29$ & 2.17 & $0.16^{+0.02}_{-0.02}$  &  $0.45^{+0.05}_{-0.06}$  & $0.98^{+0.04}_{-0.03}$  & $1.16^{+0.15}_{-0.11}$ & $-1.82^{+0.10}_{-0.10}$&$1.4^{+0.4}_{-0.3}$  &  $0.8^{+0.2}_{-0.2}$  \\
$2.29-2.45$ & 2.38 & $0.18^{+0.02}_{-0.04}$  &  $0.51^{+0.07}_{-0.12}$  & $0.97^{+0.12}_{-0.04}$  & $1.15^{+0.24}_{-0.12}$ & $-1.46^{+0.10}_{-0.10}$& $1.6^{+1.3}_{-0.6}$  &  $0.7^{+0.3}_{-0.2}$ \\
$2.45-2.60$ & 2.55 & $0.13^{+0.02}_{-0.03}$  &  $0.37^{+0.07}_{-0.08}$   & $1.08^{+0.03}_{-0.04}$  & $1.10^{+0.15}_{-0.08}$ & $-1.58^{+0.11}_{-0.10}$& $2.0^{+0.6}_{-0.5}$  &  $0.6^{+0.1}_{-0.1}$   \\
$2.60-2.81$ & 2.72 & $0.19^{+0.02}_{-0.04}$  &  $0.53^{+0.05}_{-0.11}$  & $0.99^{+0.05}_{-0.03}$  & $1.40^{+0.25}_{-0.21}$ & $-1.25^{+0.10}_{-0.11}$ & $2.0^{+0.5}_{-0.5}$  &  $0.6^{+0.1}_{-0.1}$  \\
$2.81-3.05$ & 2.94 & $0.13^{+0.03}_{-0.03}$  &  $0.36^{+0.10}_{-0.09}$  & $1.10^{+0.05}_{-0.05}$  & $1.31^{+0.30}_{-0.22}$ & $-1.18^{+0.10}_{-0.09}$ & $2.2^{+1.0}_{-0.7}$  &  $0.5^{+0.1}_{-0.1}$  \\
$3.05-3.73$ & 3.26 & $0.11^{+0.02}_{-0.02}$  &  $0.30^{+0.06}_{-0.07}$  &  $1.12^{+0.03}_{-0.03}$ & $1.45^{+0.25}_{-0.21}$ & $-1.14^{+0.07}_{-0.08}$& $2.1^{+0.5}_{-0.4}$  &  $0.6^{+0.1}_{-0.1}$ 
\enddata
\tablecomments{$\bar{z}$ is the mean redshift in a bin; $\gamma-1$ is derived following Equation~(\ref{eq:Gamma-gamma})
}
\end{deluxetable*}

\section{Results and discussion}\label{sec:discussion}

\begin{figure}[ht]
\includegraphics [width=\columnwidth]{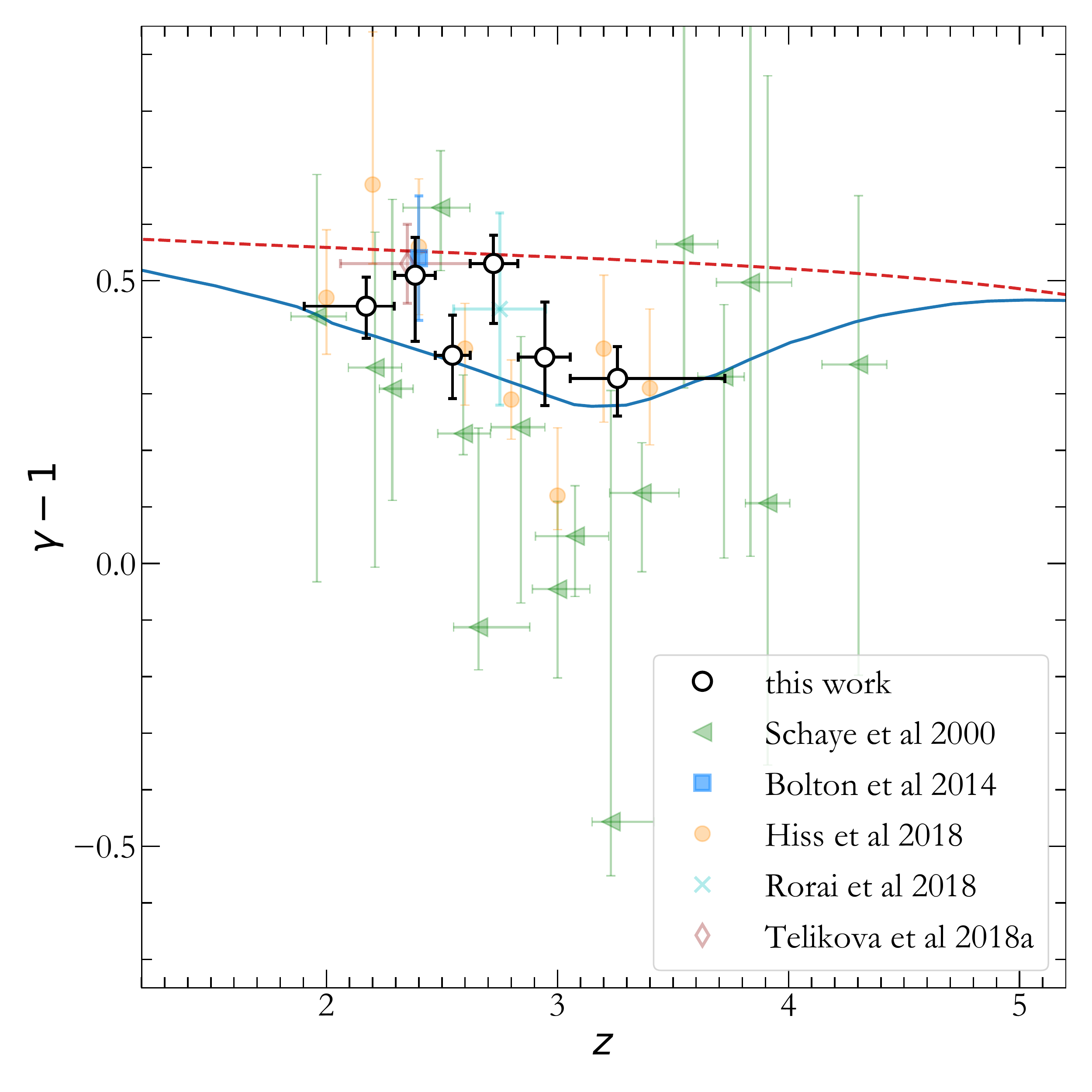}
\caption{Evolution of the power-law index $\gamma$ with redshift $z$. Solid and dash-dotted lines correspond to models from \citet{UptonSanderbeck2016} with and without account for \HeII\ reionization, respectively.}
\label{fig:gamma}
\end{figure}
\begin{figure}[ht]

\includegraphics [width=\columnwidth]{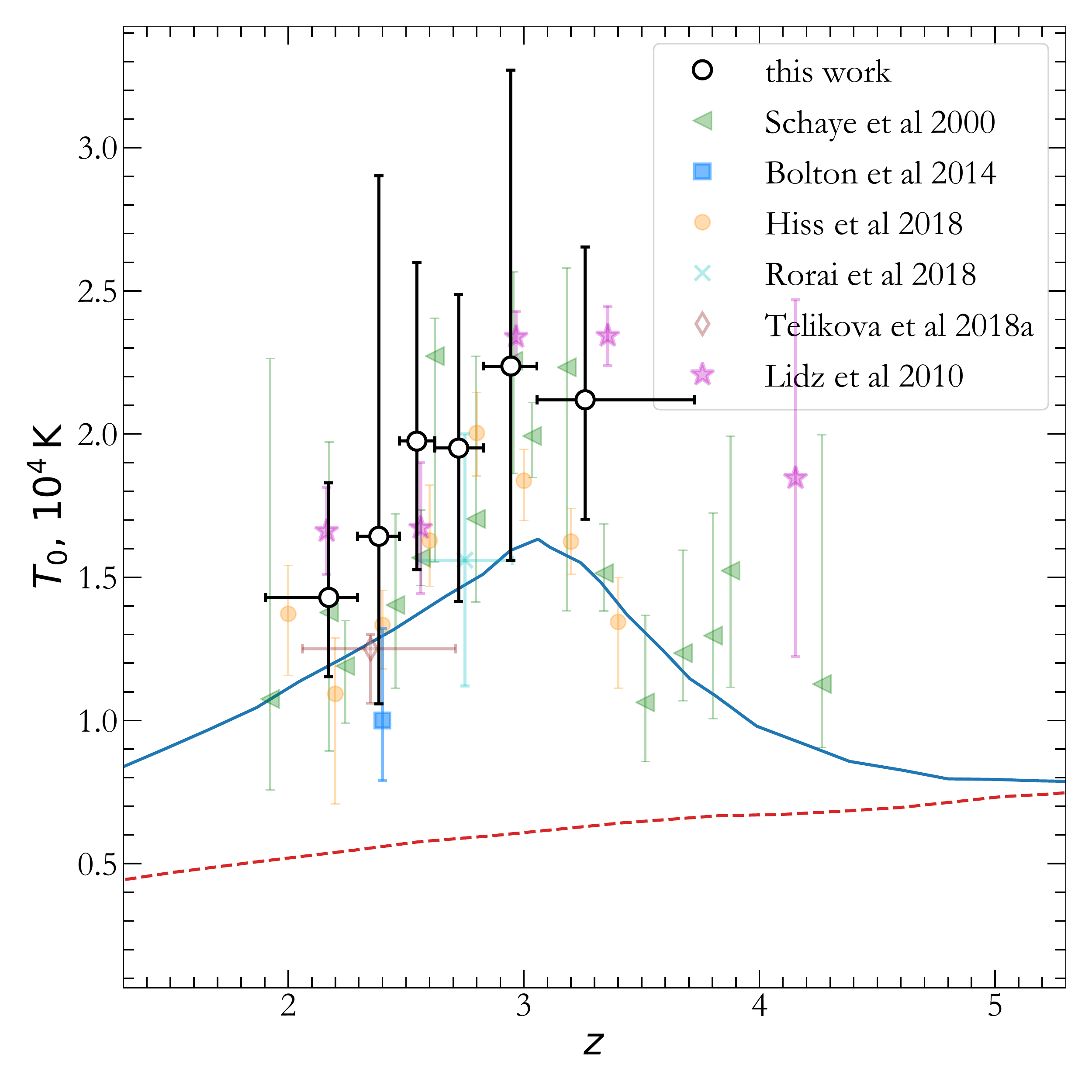}

\caption{Evolution of the temperature at the mean density $T_0$ with redshift $z$. Solid and dash-dotted lines correspond to models from \citet{UptonSanderbeck2016} with and without account for \HeII\ reionization, respectively.}
\label{fig:T0}
\end{figure}

\begin{figure}[ht]
\centering
\includegraphics [width=\columnwidth]{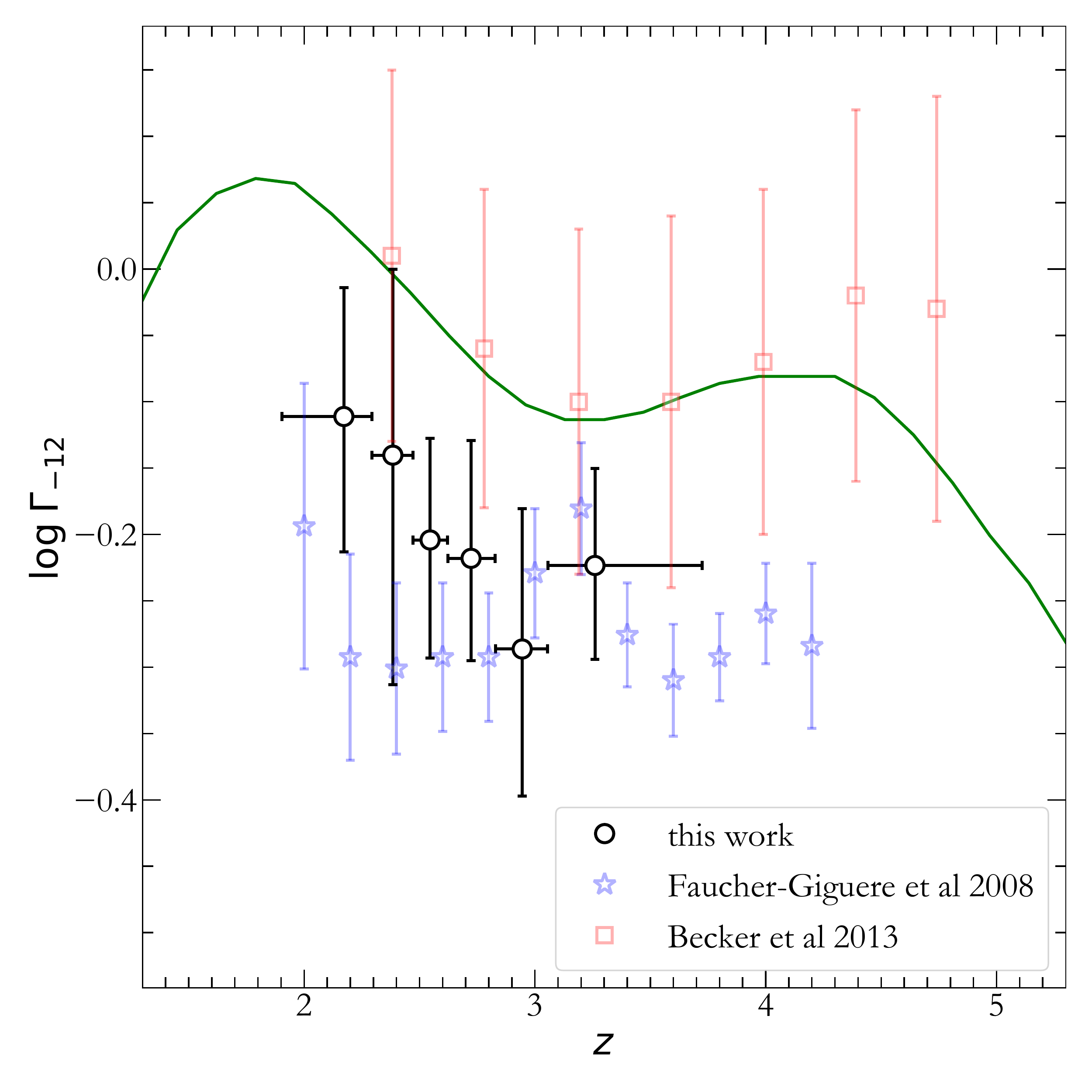}
\caption{Evolution of the hydrogen photoionization rate $\Gamma_{-12}$ with redshift $z$. Measurements by \cite{Faucher2008} and \cite{Becker2013MNRAS} are shown by stars and open squares and assume $\gamma-1 = 0.63$ and $\gamma-1 = 0.40$ respectively. Our $\Gamma_{-12}$ values (open circles) are inferred from adaptive $\gamma(z)$ from Table~\ref{tab:pars}. Notice, that in this figure measurements by \cite{Becker2013MNRAS}  and  \cite{Faucher2008} are not corrected to the cosmology used in this paper. The solid curve shows a model from \citep{Khaire2019}.}
\label{fig:G12}
\end{figure}

From the obtained posterior distributions of the model parameters 
we can directly infer the {TDR} parameter $\gamma(z)$  using Equation~\eqref{eq:Gamma-gamma}. However, the {TDR} in Equation~(\ref{eq:eos}) is also determined by $T_0(z)$, which can not be directly inferred from the model parameters. As discussed in Section~\ref{sec:igm_eos}, we can estimate $\Gamma$ and $T_0$ individually by combining our formalism with the measurements of the Ly$\alpha$ forest opacity. {Corresponding values of $\Gamma$ and $T_0$ are given in Table~\ref{tab:pars}. Notice, that the uncertainties of these parameters, shown in Table~\ref{tab:pars} and Figures~\ref{fig:T0} and \ref{fig:G12}, include the measurement errors of the effective optical depth from \citet{Faucher2008tau}.}

Evolution of the power-law index of the IGM {TDR} is shown in Figure~\ref{fig:gamma}. Our measurements of $\gamma$ are in a good agreement with the results of \cite{Hiss2017,Schaye2000}. Notice, that our measurements have smaller uncertainties than that in the previous works, however, these uncertainties are purely statistical and the systematic uncertainties are not estimated here. The power-law index $\gamma(z)$ apparently decreases with increasing redshift and reaches $\gamma-1 \approx 0.3$ at $z\approx3$. However, the linear regression slope is $a=-0.11\pm0.07$, that means that this decrease is not statistically significant. Nevertheless, in four bins (including two with highest redshifts) we see that $\gamma-1$ is at least 2$\sigma$ lower than the asymptotic value of 0.6. This decrease can be explained by the presence of the \HeII\ reionization event at $z\gtrsim 3$. We illustrate this in Figures~\ref{fig:gamma} and \ref{fig:T0} by plotting the {TDR} parameters evolution simulated by \citet{UptonSanderbeck2016} with account for \HeII\ reionization (blue solid lines) and without \HeII\ reionization (red dashed lines). Clearly, the solid curve agrees with the data much better than the dashed curve. Notice, however, that these simulations are not fully independent on the observational data, since \citet{UptonSanderbeck2016} used the measurements of the temperature by \cite{Becker2011} and \citet{Boera2014} and adjusted the simulation results to them. 
Additionally, $T_0(z)$ (Figure~\ref{fig:T0}) reaches the temperatures up to $(2-3)\times10^4$~K, also consistent with \HeII\ reionization event \citep{McQuinn2009ApJ, UptonSanderbeck2016}. However, the uncertainties for $T_0(z)$ are quite large, and we can not make conclusive statement about the character of its evolution with $z$. In summary, our results are consistent with the scenario of \HeII\ reionization at $z\gtrsim3$.

Comparison of the $\Gamma$ parameter estimated in this work with the results by \cite{Faucher2008} and \cite{Becker2013MNRAS} is shown in Figure~\ref{fig:G12}. We do not correct measurements by \cite{Becker2013MNRAS} for the different cosmology used by them, but this effect is negligible in contrast to difference in $\gamma$ and $T_0$. 
Notice, that measurements from \cite{Becker2013MNRAS}  and \cite{Faucher2008} shown in  Figure~\ref{fig:G12} correspond to the  {TDR} slopes $\gamma-1=0.40$ and $\gamma-1=0.63$, respectively.
The main reason of disagreement between these measurements of $\Gamma$ is a different temperature $T_0$. Temperature at the mean density inferred in our work is lower than the temperature used in \citep{Faucher2008} and higher than that in \cite{Becker2013MNRAS}. Consequently, our measurements  of $\Gamma$ are located between those by \cite{Becker2013MNRAS} and \cite{Faucher2008}. Likewise the results of \cite{Faucher2008} and in contrast to the results of \cite{Becker2013MNRAS}, our measurements are systematically lower than values of $\Gamma$ derived from the recent model of UV background \cite{Khaire2019}.

The results discussed above are derived under assumption that the characteristic size of the absorber cloud is equal to the Jeans length, see Equation~(\ref{eq:rho-N}) and \cite{Schaye2001}. In fact, these scales might be close, but not necessarily equal \citep{Schaye2001, Garzilli2015}. This, in turn, can result in biased measurements of the parameters $T_0$ and $\Gamma$. We checked, that changing the prefactor in Equation~(\ref{eq:rho-N}) by a factor of two leads to a 30\% change in $T_0$ and 20\% change in $\Gamma$. 
Moreover, our model assumes that the minimal broadening at fixed $N$ is caused only by the thermal motions in the absorber cloud. It was shown that the spatial structure of the absorbers may impact the minimal 
broadening due to the Hubble expansion \citep{Garzilli2015,Garzilli2018}. This 
may lead to the systematic underestimation of the {TDR} power-law index $\gamma$ and can also impact the $T_0$ and $\Gamma$ measurements. However, one expects that the Hubble broadening is most important at low $N$, and its effect on the results presented here is not very strong \citep{Telikova2019}.
We plan to thoroughly address this issue in future work.

The power-law index of the column density distribution $\beta$ in Equation~(\ref{eq:L}) increases with decreasing redshift, see Table~\ref{tab:pars}. Our results are consistent with the results from \cite{Kim2002} who give $\beta=-1.42\pm0.03$ for the redshift range $z=1.53-2.57$ and $\beta=-1.11\pm0.04$ for the redshift range $z=2.96-3.60$ 
and column densities  $\log N\  [\mathrm{cm}^{-2}] = 12.5-14$. 

Our results were obtained using the parametric model for the $(N,\, b)$ distribution.
Although the selected shape fits the data well, the choice of the model for the distribution can impose some systematic uncertainties. 
Recently, \citet{Hiss2019} developed a method for comparison of the observed $(N,\, b)$ distribution with model distributions resulting from the analysis of the  mock Ly$\alpha$ forest  generated based on the hydrodynamical simulations of the IGM. The simulations are carried out on a grid of the IGM thermal parameters ($T_0$ and $\gamma$) and extended to the inter-grid values by means of the Gaussian process interpolation of the principal components of the kernel-density smoothed $(N,\, b)$ maps. This method is similar to various cutoff-fitting methods where the position of the cutoff (or other discrete measures, such as, e.g., a distribution median \citep{Rorai2018}) is calibrated trough the hydrodynamical simulations, however \citet{Hiss2019} calibrate the whole lines parameters distribution instead. The main complication with this approach is that the simulations are actually computationally expensive, especially when the line sample size is large, so that the ($T_0$, $\gamma$) grid resolution need to be small enough to obtain the robust inference of the thermal parameters which is free from the interpolation errors \citep{Hiss2019}. 

{While in our approach, where estimation of the parameter $\gamma$ (as well as $\beta$ and $p$) does not rely on the simulations, the main source of the systematic errors may come from the choice of the analytical form of the distribution, in the approach of \cite{Hiss2019} the possible systematics may come from the selection of the input to the hydrodynamical simulations and their calibration. Additionally, in our work we explicitly took into account outliers  by means of the mixture model for the obtained distribution, instead of the more common 2$\sigma$ rejection procedure. 
However, our inference of the parameters $\Gamma$ and $T_0$ relies on simulation results by \citet{Springel2003}. Certainly, as it was declared by \citet{Bolton2005MNRAS,Faucher2008}, the gas density probability distribution depends both on the physical parameters of used simulations, in particular cosmology and UV background,  and numerical parameters, such as box size and mass resolution. All these parameters may lead to underestimation of the $T_0$ and $\Gamma$ uncertainties by our method, which are already quite large.}

\section{Conclusions}\label{sec:conclusions}
We presented a new method to estimate parameters of the IGM {thermal state} and their evolution with redshift, which is based on the analysis of the $(N,\, b)$ distribution of the Ly$\alpha$ forest lines in quasar spectra. We developed the automatic routine to analyse quasar spectra. We selected and fit only Ly$\alpha$ forest lines with a simple velocity structure. We applied our strategy to 47 high-resolution quasar spectra from KODIAQ. Then we carefully cleaned the obtained sample from metal absorption lines by visual inspection and then compared observed distribution of $(N,\, b)$ parameters in six redshift bins with the model 2D joint probability density function. We also took into account the outliers by means of the mixture model. 
As a result, we obtained the physical parameters of the IGM as a function of redshift. 

We found that the power-law index $\gamma$ in the IGM {TDR} has a tendency to decrease with increasing redshift and reaches $\gamma-1\approx0.30$ at $z\sim3$. The $\gamma(z)$ dependence is in agreement with the predictions of the evolution scenario in which \HeII\ reionization takes place near $z\gtrsim3$. 
To construct the temperature at the mean density $T_0$ and the hydrogen photoionization rate $\Gamma$, we used the model probability density function of the gas distribution proposed in \citet{Miralda-Escude2000} and measurements of the Ly$\alpha$ forest opacity by  \citet{Faucher2008tau}. Although uncertainties of $T_0(z)$ are quite large (Figure~\ref{fig:T0}), the typical values are high, reaching a maximum value $T_0\approx(2.2\pm1.0)\times 10^4$~K at $z\sim 3$. This  also favours the \HeII\ reionization scenario, however we can not make a conclusive statement about its evolution with the redshift. The same is true for the $\Gamma$ parameter (see Figure~\ref{fig:G12}). The mean value of the phtoionization rate over the redshift range considered  in our study is $\Gamma_{-12}\approx0.6$.

The power-law index  $\beta$ of the column density distribution increases with decreasing $z$ from $\beta\approx-1.14$ at $\bar{z}=3.26$ to $\beta\approx-1.82$ at $\bar{z}=2.17$, consistent with previous measurements \citep{Kim2002}.

As discussed in Section~\ref{sec:discussion}, in the present analysis we did not include the effects which accounts for the extent and exact structure of the absorbers.
We estimated that changes of characteristic absorber size by factor of 2 result in $\lesssim 30$\% systematical uncertainty for inferred parameters. Additional source of uncertainty is the broadening due to the Hubble expansion \citep{Garzilli2015}. 
Taking into account these effects can result in more comprehensive analysis of the evolution of the physical parameters of the IGM. We plan to address these issues in future work.

In the present study we used a large sample of quasar spectra from a uniform database, obtained on the same instrument and reduced with the same algorithms. Nevertheless, the statistical uncertainties are quite large. To reduce the uncertainties one needs to increase the number of the quasar spectra.
This can be provided either using the archival data from other instruments (such as VLT/UVES) or with the advent of the next generation of the optical telescopes, like the Extremely Large Telescope and Thirty Meter Telescope.  

\acknowledgments{
The work was supported by the Russian Science Foundation grant 18-72-00110. 
}

\vspace{5mm}
\facilities{KECK}

\software{AstroPy \citep{Astropy2013A&A,Astropy2018AJ},\,  SciPy \citep{Scipy},\, Matplotlib \citep{Hunter:2007},\, ChainConsumer \citep{Hinton2016}.}

\appendix

\section{Observational data}\label{sec:app_data}

\startlongtable
\begin{deluxetable*}{*{5}{C}}
    \tablecaption{Quasar spectra\label{tab:qsos}}
    \tablehead{
    \colhead{Quasar} & \colhead{$z_{\rm em}$\tablenotemark{a}} &\colhead{$z_{\rm range}$\tablenotemark{b}} & \colhead{C/N} & \colhead{Resolution}
} 
\startdata
J020950-000506&2.83 &$2.246-2.792$&  66&    48000\\
J155152+191104&2.84 &$2.255-2.802$& 111&    36000\\ 
J162548+264658&2.52 &$1.983-2.486$&  32&    36000\\
J170100+641209&2.73 &$2.161-2.694$&  67&    72000\\	
J010311+131617&2.72 &$2.153-2.684$&  99&    36000\\
J082107+310751&2.63 &$2.076-2.595$&  49&    48000\\
J014516-094517&2.72 &$2.153-2.684$&  32&    48000\\
J101155+294141&2.65 &$2.093-2.615$& 117&    48000\\
J220852-194400&2.57 &$2.025-2.536$&  76&    36000\\
J082619+314848&3.09 &$2.467-3.049$&	 30&    48000\\	
J121117+042222&2.53 &$1.991-2.496$&	 32&    48000\\	
J101723-204658&2.54 &$2.000-2.506$&	 53&	48000\\	
J234628+124859&2.57 &$2.025-2.536$&	 69&	48000\\
J121930+494052&2.63 &$2.076-2.595$&	 39&	48000\\	
J143500+535953&2.63 &$2.076-2.595$&	 54&	48000\\	
J081240+320808&2.71 &$2.144-2.674$&	 37&	48000\\
J012156+144823&2.87 &$2.280-2.832$&	 42&	48000\\	
J143316+313126&2.94 &$2.340-2.901$&	 55&	36000\\
J120917+113830&3.10 &$2.476-3.059$&	 50&	36000\\	
J102009+104002&3.17 &$2.535-3.128$&	 39&	36000\\	
J095852+120245&3.30 &$2.646-3.256$&	 49&	36000\\
J173352+540030&3.42 &$2.747-3.374$&	 49&	36000\\
J193957-100241&3.79 &$3.062-3.739$&	 76&	48000\\
J003501-091817&2.42 &$1.898-2.388$&	 34&	36000\\
J004530-261709&3.44 &$2.764-3.394$&	 23&	48000\\
J010806+163550&2.65 &$2.093-2.615$&	 56&	36000\\
J010925-210257&3.23 &$2.586-3.187$&	 19&	36000\\
J030341-002321&3.18 &$2.544-3.138$&	 26&	48000\\
J045213-164012&2.68 &$2.119-2.644$&	 72&	36000\\
J073149+285448&3.68 &$2.968-3.631$&	 28&	36000\\
J074521+473436&3.22 &$2.578-3.177$&	 42&	48000\\
J083102+335803&2.43 &$1.907-2.398$&	 32&	36000\\
J092914+282529&3.40 &$2.731-3.355$&	 61&	72000\\
J094202+042244&3.27 &$2.620-3.226$&	 66&	48000\\
J100841+362319&3.13 &$2.501-3.088$&	 35&	48000\\
J101447+430030&3.12 &$2.493-3.078$&	 61&	36000\\
J102325+514251&3.45 &$2.773-3.404$&	 22&	48000\\
J113130+604420&2.92 &$2.323-2.881$&	 23&	48000\\
J113418+574204&3.52 &$2.832-3.473$&	 32&	36000\\
J113508+222715&2.89 &$2.297-2.851$&	 28&	48000\\
J122518+483116&3.09 &$2.467-3.049$&	 33&	48000\\
J124610+303131&2.56 &$2.017-2.526$&	 26&	48000\\
J130411+295348&2.85 &$2.263-2.812$&	 34&	48000\\
J135038-251216&2.53 &$1.992-2.496$&	 42&	48000\\
J144453+291905&2.66 &$1.992-2.624$&	101&    48000\\
J155810-003120&2.83 &$2.246-2.792$&	 70&	36000\\
J160455+381214&2.55 &$2.009-2.516$&	116&    36000
\enddata
\tablenotetext{a}{quasar redshift}
\tablenotetext{b}{redshift range used for Ly$\alpha$ forest analysis}
\end{deluxetable*}

\newpage\clearpage
\onecolumngrid
\section{Posterior distributions}\label{sect:posteriors}
\begin{figure*}[h!]
 \subfloat[$\bar{z}=2.17$]{\includegraphics[width = 0.33\textwidth]{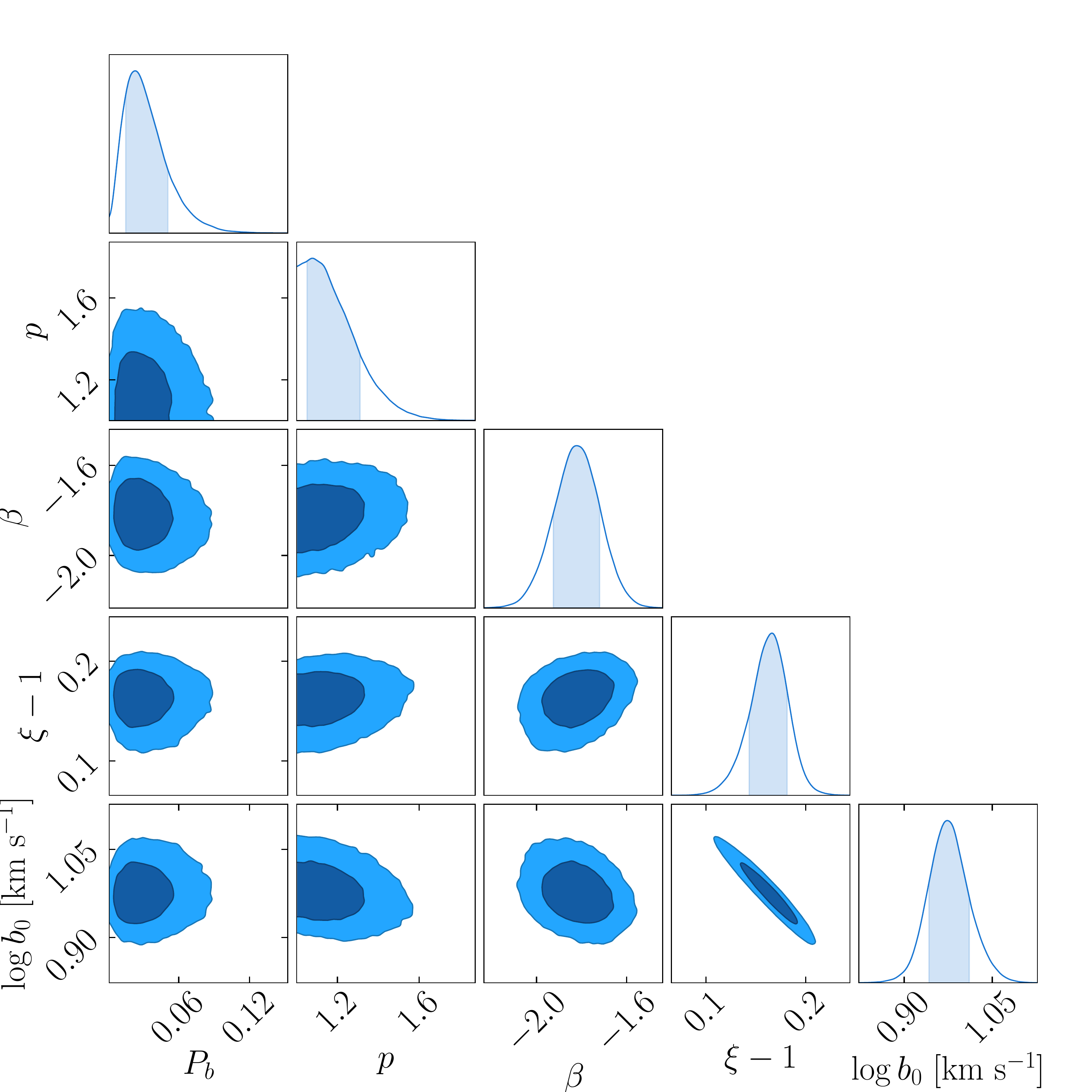}} 
\subfloat[$\bar{z}=2.38$]{\includegraphics[width = 0.33\textwidth]{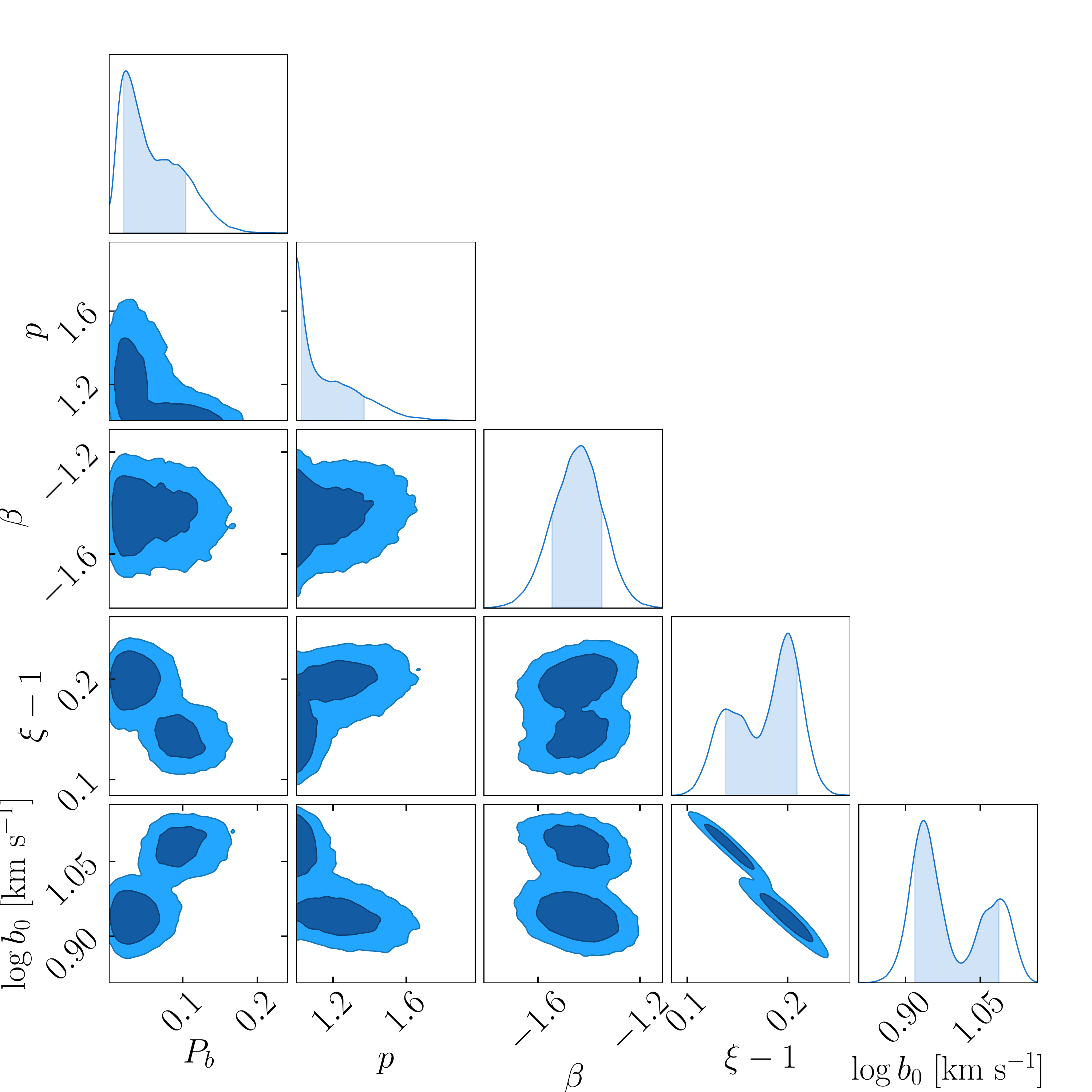}}
\subfloat[$\bar{z}=2.55$]{\includegraphics[width = 0.33\textwidth]{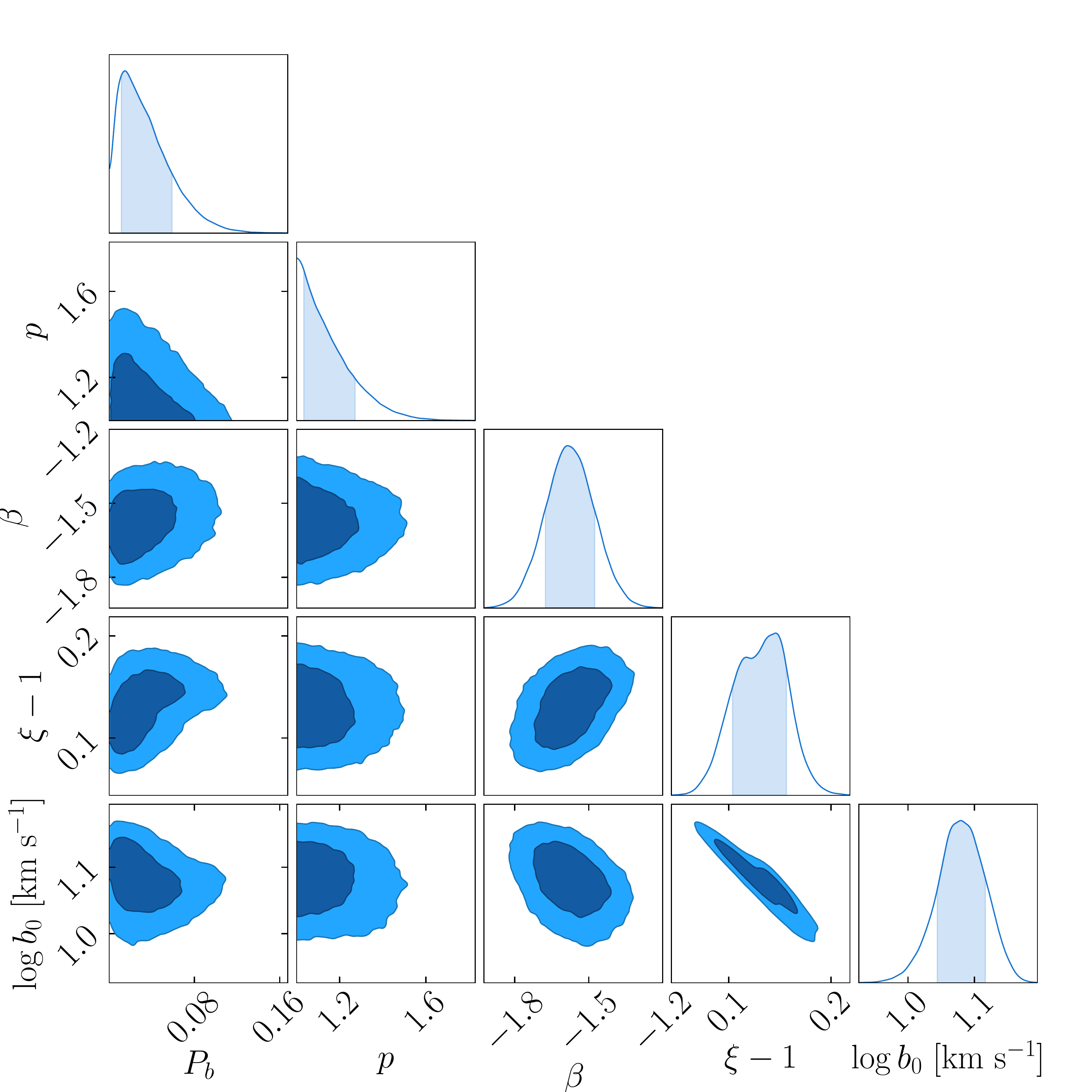}}\\
\subfloat[$\bar{z}=2.72$]{\includegraphics[width = 0.33\textwidth]{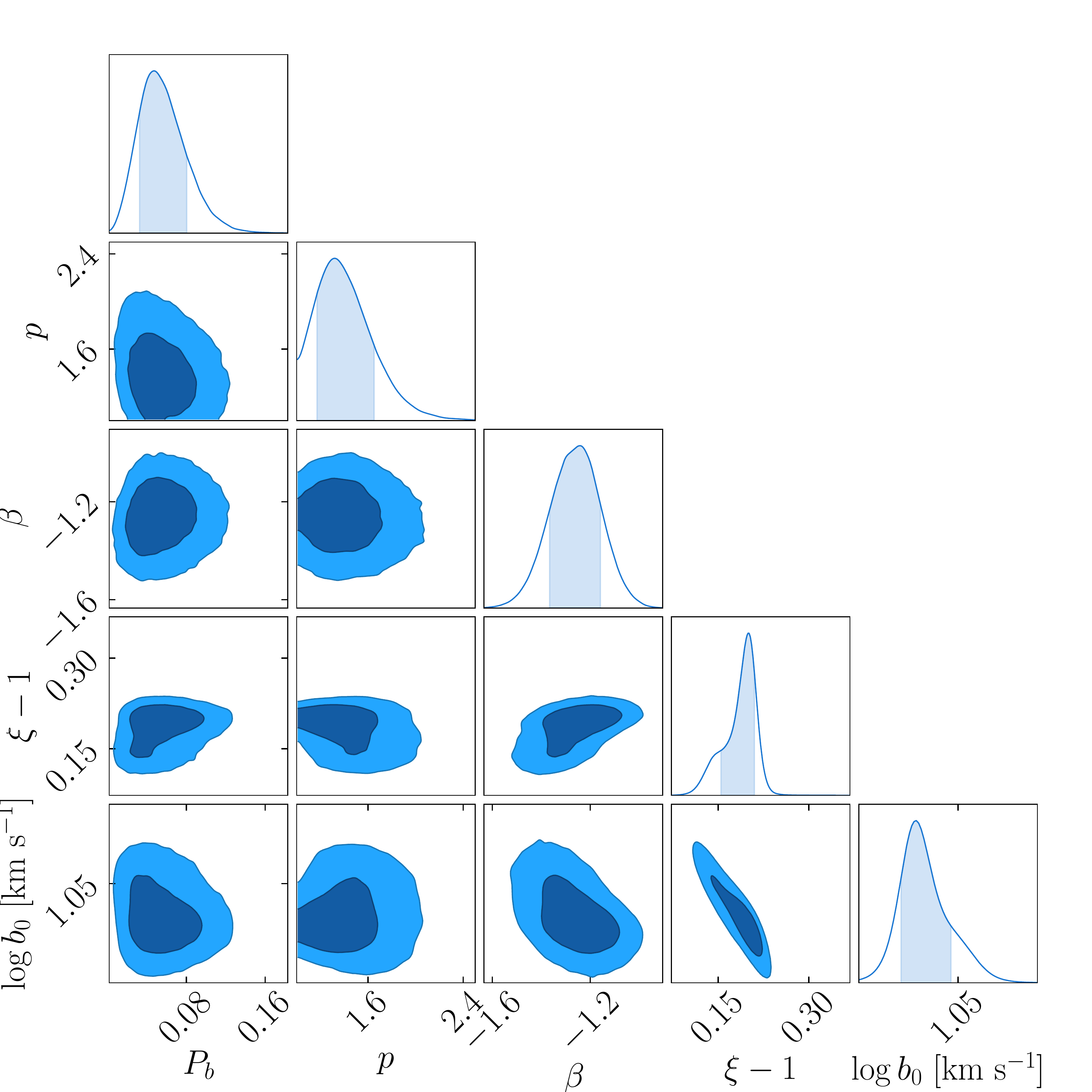}}
\subfloat[$\bar{z}=2.94$]{\includegraphics[width = 0.33\textwidth]{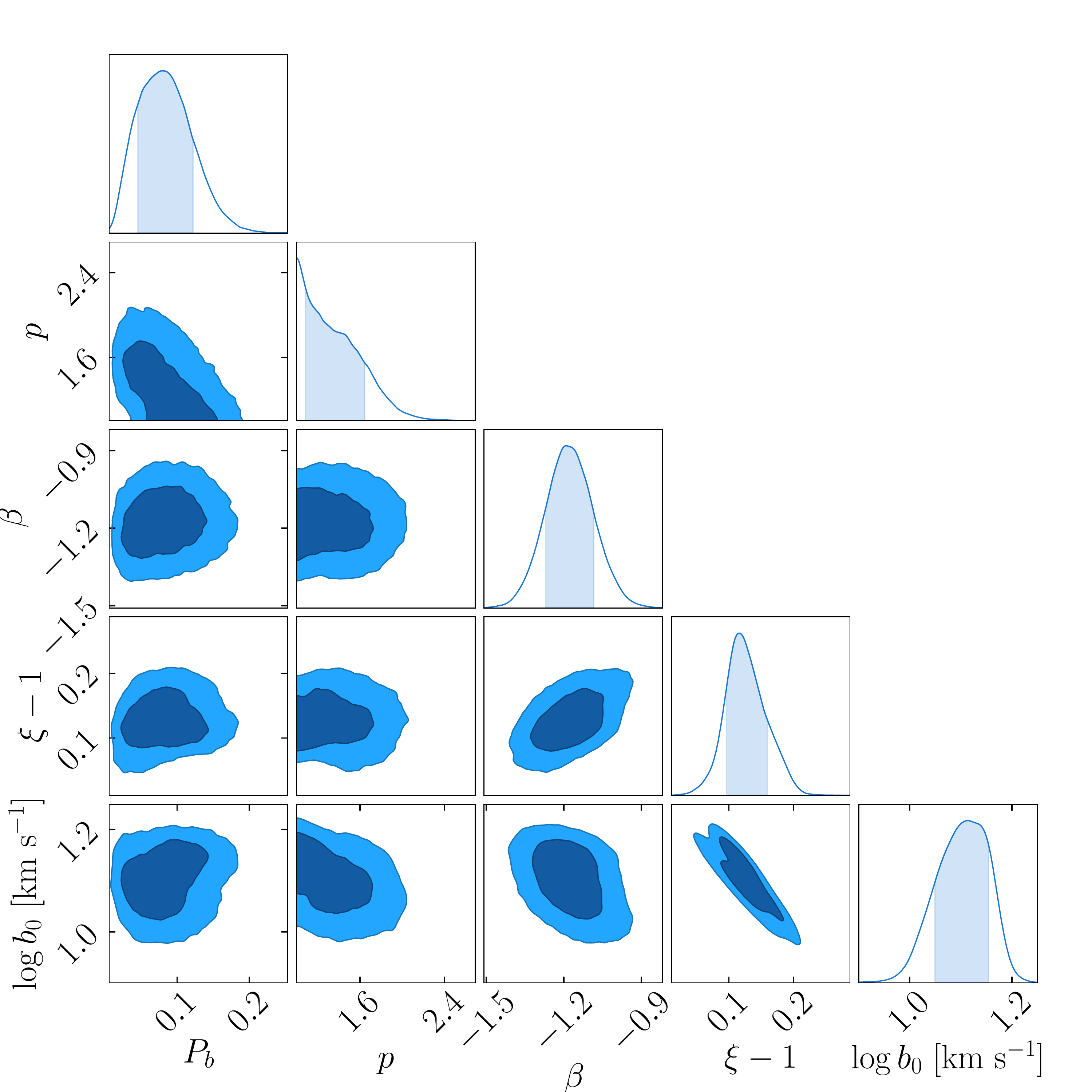}}
\subfloat[$\bar{z}=3.26$]{\includegraphics[width = 0.33\textwidth]{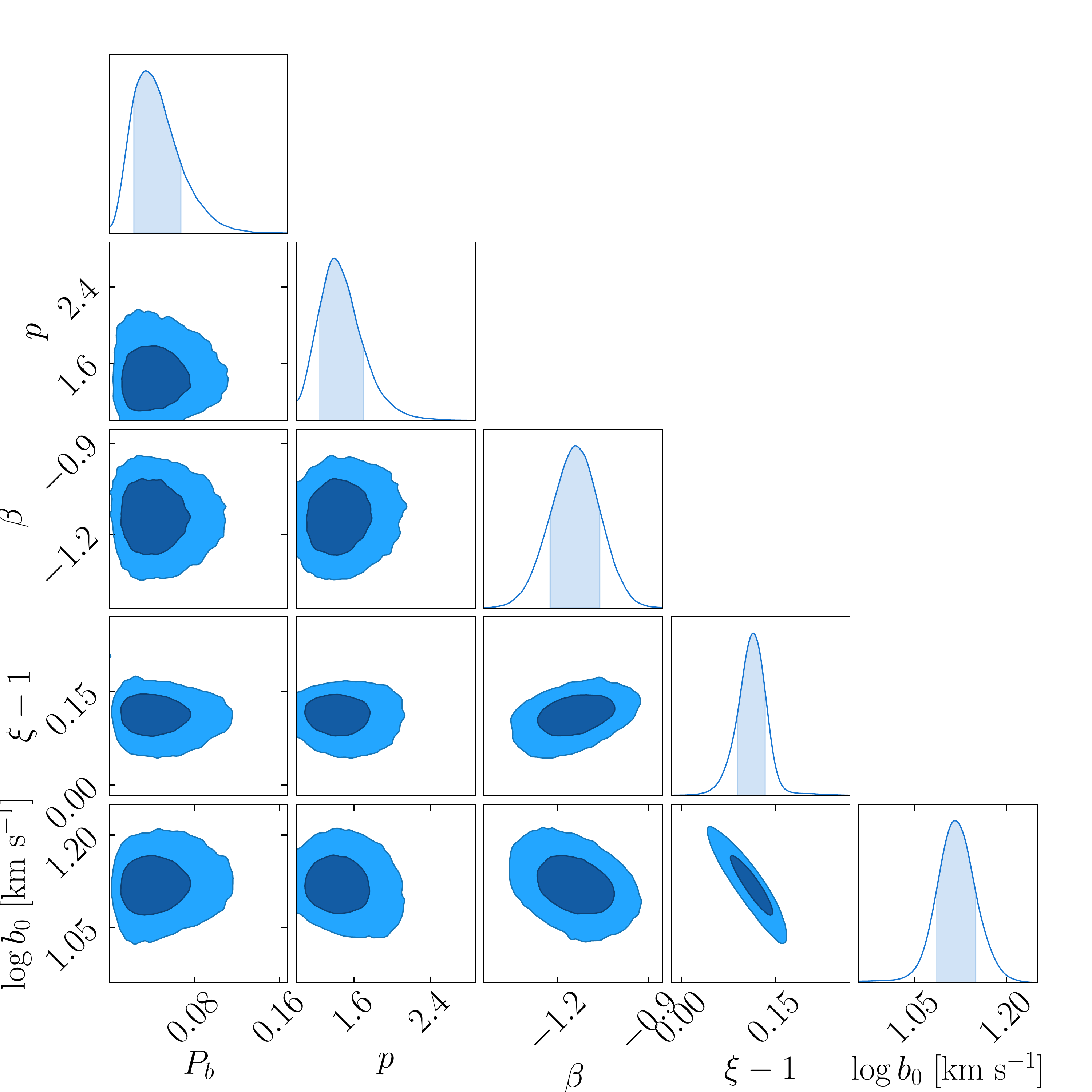}}
\caption{Posterior distributions of the estimated parameters for 6 redshift bins. Dark and light filled areas correspond to the 68\% and 95\% credible regions, respectively.}\label{fig:triangles}
\end{figure*}

{Figure~\ref{fig:triangles} shows the 1D and 2D marginalized posterior distributions of the model parameters. Each panel corresponds to one redshift bin with mean redshift values indicated at the panel captions. In most cases we obtained simple bell-like shaped 1D posteriors, which allow us to provide reasonable interval estimates of the parameters. We note that in the second bin we obtain a bimodal distribution with two separated regions seen in 1D as well as 2D plots. 
We checked that this feature does not result from a problem with the convergence of the MCMC chain for this bin, rather this bimodality reflects the underlying data which require a prominent drop in the value of $\xi-1$ (and $\gamma-1$) parameter in the vicinity of the second redshift bin, see Figure~\ref{fig:gamma} and Table~\ref{tab:pars}. In order to check this, we considered two shifted bins with  $2.17<z< 2.38$ and $2.38<z<2.55$. These bins contain the same number of lines as our original bins, but their edges are shifted by half the bin size. After performing the fitting, we got $\xi-1=0.19\pm0.03$ and $\log\, b_0 \:[\mathrm{km}~\mathrm{s}^{-1}]=0.92\pm0.04$ for the first shifted bin and $\xi-1=0.11\pm0.02$ and $\log\, b_0 \:[\mathrm{km}~\mathrm{s}^{-1}]=1.10\pm0.02$ for the second shifted bin, the values close to the positions of the modes in the bimodal distribution in Figure~\ref{fig:triangles}(b). It is interesting to investigate this region with higher resolution in $z$, which would be possible with increase of the sample size. }

\twocolumngrid

\section{Fisher matrix for Voigt profile}
\label{sect:fisher}
Here we provide the equations used to calculate the Fisher matrix for the single absorption line with Voigt profile. 
The Fisher matrix formalism is the standard way to forecast the experiment uncertainties, based on the known likelihood function. Here we use uncertainties $\sigma_N$ and $\sigma_b$ of parameters $N$ and $b$ calculated from the Fisher matrix to estimate the step sizes of the grid, which was used for the searching Ly$\alpha$ forest lines.
We assume that the redshift of the line is fixed (i.e., it is not the parameter of the fit), hence we provide the equations for these two Voigt profile parameters.

The components of the Fisher matrix for two parameters are 

\begin{equation}\label{eq:Fisher}
	F_{kl} = \mathbb{E}\left(\frac{\partial^2 \ln{\mathcal L}}{\partial \theta_k\partial \theta_l}\right)  
\approx 2 (C/N)^2 \sum\limits_{i=1}^{n}\frac{\partial I_{\rm conv}}{\partial \theta_k}\frac{\partial I_{\rm conv}}{\partial \theta_l},
\end{equation}
where $\theta_k$ is $b$ or $N$ and mathematical expectation is taken over the probability distribution function in each pixel. We also roughly suggested that pixels have the same uncertainties within the line, $\sigma$, i.e. $C/N \approx 1/\sigma$. The partial derivative of the convolved profiles $I_{\rm conv}$ (given in Equation~\ref{eq:convol}) is
\begin{equation}
\frac{\partial I_{\rm conv}}{\partial \theta_k} = G \otimes\frac{\partial I}{\partial \theta_k}  = G \otimes \left(e^{-\tau} \frac{\partial \tau}{\partial \theta_k}\right),
\end{equation}
where $\otimes$ stands for the convolution, and the partial derivatives of optical depth, $\tau$, (given in Equation~\ref{eq:tau}) over $b$ and $N$ are
\begin{eqnarray}
	\frac{\partial \tau}{\partial N}  &=& \frac{\tau_0 H}{N}, \\
    \frac{\partial \tau}{\partial b} &=& -\frac{\tau}{b} + \tau_0\left(\frac{\partial H(a,\,x)}{\partial x} \frac{\partial x}{\partial b} + \frac{\partial H(a,\,x)}{\partial a} \frac{\partial a}{\partial b}\right) \nonumber\\
	&=& -\frac{2\tau_0}{b} \left(H(a,\,x)\left(a^2-x^2+\frac{1}{2}\right) \right.\nonumber\\ 
	&+&\left.2K(a,\,x)\:ax - \frac{a}{\sqrt{\pi}}\right),\label{eq:dtau_db}
\end{eqnarray}
where $H(a,\,x)$ is the Voigt function of variables $a$ and $x$ (see Equation~(\ref{eq:voigt})) and $K(a,\,x)$ is the imaginary Voigt function \citep{Heinzel1978}.

To apply Equations~(\ref{eq:Fisher})--(\ref{eq:dtau_db}), we need to specify certain signal to noise ratio and the full width at the half maximum (FWHM) of instrumental function $G$ (we assume $G$ to be Gaussian). 
Given the resolution of the spectrum $R$, we set $\rm FWHM = \lambda / R$, where $\lambda$ is the median wavelength in the searched region. We also assume the pixel size to be FWHM/3, which is typical for the observed optical spectra. 

Once the Fisher matrix for given $b$ and $N$ is calculated, its inverse gives the covariance matrix for these parameters. The square roots of the diagonal elements of the covariance matrix were used further to estimate the uncertainties, which determines the steps of the adaptive grid in $(N,\, b)$ parameter space, used during the search, see Section~\ref{sec:fitting procedure}.

\section{Probability density distribution function}\label{sect:pdf_var_parameters}
\begin{figure*}[ht]
\centering
\includegraphics [width=0.95\textwidth]{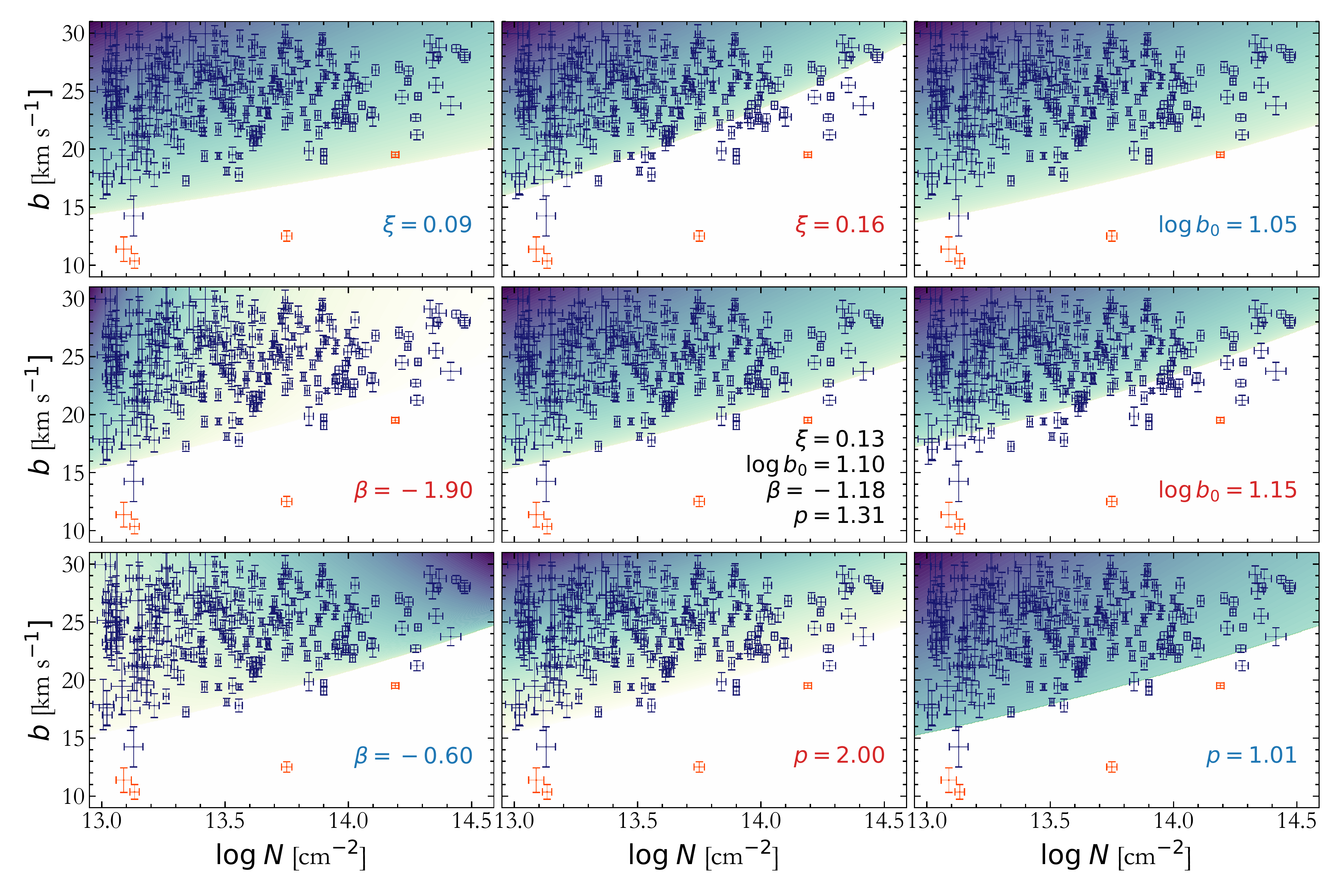}
\caption{Blue error crosses show the Ly$\alpha$ lines in the redshift range $z=2.81-3.05$ after the metal lines rejection (fifth redshift bin in Figure~\ref{fig:data_pdf}).  
The outliers are shown by the red error crosses.
The best-fit distribution function is shown by the colour gradient area in the central panel. 
Distribution functions, calculated by varying value one of the model parameters $\xi,\,\log b_0,\, p$, and $\beta$ while keeping other parameters fixed, are shown in the other panels. Values of the varied parameters are shown in the each panel. Blue (red) label colours corresponds to the values, which are less (greater) than that for the best-fit distribution.
}
\label{fig:data_pdf_var_par}
\end{figure*}
{In Figure~\ref{fig:data_pdf_var_par} for illustration purposes we show how the probability density function in Equation~(\ref{eq:pdf_N_b}) changes with variation of its main parameters. The fifth redshift bin $z=2.81-3.05$ is chosen as an example. }

\bibliographystyle{aasjournal}


\end{document}